\documentclass[]{pasj02} 

\usepackage[dvipsnames]{xcolor}
\usepackage{natbib} 
\definecolor{xlinkcolor}{cmyk}{1,1,0,0}
\usepackage{url}
\usepackage{comment}
\usepackage{lscape}
\usepackage{multirow}
\usepackage{threeparttable}
\usepackage{bm}

\usepackage[colorlinks=true, linkcolor=xlinkcolor, citecolor=xlinkcolor, urlcolor=xlinkcolor]{hyperref}
\makeatletter
\providecommand{\theH@article}{}

\makeatother

\newcommand{\Hi}{H\,{\sc i}}

\newcommand{\Nii}{N\,{\sc ii}}

\newcommand{\Oiii}{[O\,{\sc iii}]}

\newcommand{\taudep}{\tau_\mathrm{dep}}
\newcommand{\mugas}{\mu_\mathrm{gas}}
\newcommand{\Mgas}{M_\mathrm{gas}}

\newcommand{\Mdust}{M_\mathrm{dust}}
\newcommand{\Umean}{\langle U \rangle}
\newcommand{\Tdust}{T_\mathrm{dust}}

\newcommand{\SFR}{\mathrm{SFR}}
\newcommand{\sSFR}{\mathrm{sSFR}}
\newcommand{\SFE}{\mathrm{SFE}}
\newcommand{\LIR}{L_\mathrm{IR}}
\newcommand{\dMS}{\Delta_\mathrm{MS}}

\newcommand{\gdr}{\delta_\mathrm{GDR}}

\newcommand{\MS}[1]{\langle #1 \rangle_\mathrm{MS}}

\jyear{2026}
\Received{}
\Accepted{}

\graphicspath{{figs/}} 

\begin{document} 

\title{Gas Fraction and Depletion Time Drive the Main-Sequence Scatter in Massive Galaxies at $z\sim1.5$}

\author{
 Daichi \textsc{Kashino},\altaffilmark{1}\altemailmark\orcid{0000-0001-9044-1747} \email{daichi.kashino@nao.ac.jp; kashinod.astro@gmail.com} 
 Boris S. \textsc{Kalita},\altaffilmark{2,3}\orcid{0000-0001-9215-7053}
 John D. \textsc{Silverman},\altaffilmark{3,4,5}\orcid{0000-0002-0000-6977}
 Tomoko L. \textsc{Suzuki},\altaffilmark{3}\orcid{0000-0002-3560-1346}
 Annagrazia \textsc{Puglisi},\altaffilmark{6}\orcid{0000-0001-9369-1805}
 Alvio \textsc{Renzini},\altaffilmark{7}\orcid{0000-0002-7093-7355} 
 Emanuele \textsc{Daddi},\altaffilmark{8}\orcid{0000-0002-3331-9590}
 Giulia \textsc{Rodighiero},\altaffilmark{7,9}\orcid{0000-0002-9415-2296}
 Francesco \textsc{Sinigaglia},\altaffilmark{10}\orcid{0000-0002-0639-8043}
 Xuheng \textsc{Ding},\altaffilmark{3,11}\orcid{0000-0001-8917-2148}
 David B. \textsc{Sanders}\altaffilmark{12}\orcid{0000-0002-1233-9998}
\altaffiltext{1}{National Astronomical Observatory of Japan, 2-21-1 Osawa, Mitaka, Tokyo 181-8588, Japan}
\altaffiltext{2}{Kavli Institute for Astronomy and Astrophysics, Peking University, Beijing 100871, People's Republic of China}
\altaffiltext{3}{Kavli Institute for the Physics and Mathematics of the Universe (Kavli IPMU, WPI), UTIAS, Tokyo Institutes for Advanced Study, University of Tokyo, Chiba, 277-8583, Japan}
\altaffiltext{4}{Department of Astronomy, School of Science, The University of Tokyo, 7-3-1 Hongo, Bunkyo, Tokyo 113-0033, Japan}
\altaffiltext{5}{Center for Astrophysical Sciences, Department of Physics and Astronomy, Johns Hopkins University, Baltimore, MD 21218, USA}
\altaffiltext{6}{School of Physics and Astronomy, University of Southampton, Highfield, SO17 1BJ, UK}
\altaffiltext{7}{INAF -- Osservatorio Astronomico di Padova, Vicolo dell'Osservatorio 5, I-35122 Padova, Italy}
\altaffiltext{8}{CEA, IRFU, DAp, AIM, Universit{\'e} Paris-Saclay, Universit{\'e} Paris Cit{\'e}, Sorbonne Paris Cit{\'e}, CNRS, 91191 Gif-sur-Yvette, France}
\altaffiltext{9}{Dipartimento di Fisica e Astronomia, Universit{\`a} di Padova, Vicolo dell'Osservatorio, 3, I-35122 Padova, Italy}
\altaffiltext{10}{Institute for Fundamental Physics of the Universe (IFPU), Via Beirut 2, I-34151 Trieste, Italy}
\altaffiltext{11}{School of Physics and Technology, Wuhan University, Wuhan 430072, People's Republic of China}
\altaffiltext{12}{Institute for Astronomy, University of Hawaii, 2680 Woodlawn Drive, Honolulu, HI 96822, USA}
}



\maketitle

\begin{abstract}
We present ALMA Band 7 dust continuum observations of 57 massive ($M_\ast \gtrsim 10^{10.8}~M_\odot$) star-forming galaxies at $1.45<z<1.70$, selected from the FMOS-COSMOS survey to provide a homogeneous sample near the main sequence (MS) at cosmic noon.
The observations are sufficiently deep to yield $>3\sigma$ detections for 55 galaxies.  Combining the ALMA data with multiwavelength photometry, we reliably derive dust masses and infer molecular gas masses using metallicity-dependent gas-to-dust ratios estimated from individual metallicity measurements. The derived molecular gas mass ratio spans $\mugas = \Mgas/M_\ast=0.11\text{--}2.8$, with a median value of 0.65, corresponding to gas reservoirs more than an order of magnitude larger than in local galaxies at fixed stellar mass.  The integrated Schmidt--Kennicutt relation is consistent with previous measurements over $z=0\text{--}2$.  Across the MS, both molecular gas mass ratio and star formation efficiency scale approximately as $(\sSFR/\sSFR_\mathrm{MS})^{0.5}$, indicating that the MS scatter is driven nearly equally by variations in gas content and depletion time.  The intrinsic scatter of $0.19$~dex suggests additional galaxy-to-galaxy diversity in star formation efficiency.  Our results provide a controlled test of the unified gas scaling framework in the massive regime at $z\sim1.5$, demonstrating that the fundamental regulation of star formation through coupled modulation of gas supply and efficiency is already in place at cosmic noon.
\end{abstract}


\section{Introduction}

Understanding how galaxies build up their stellar mass and what regulates the enhancement, suppression, or quenching of star formation is central to galaxy evolution.  The advent of JWST has opened a new window onto the internal structures of galaxies at cosmic noon ($z\sim1$--3) and beyond \cite[e.g.,][]{2022ApJ...938L..24F,2023ApJ...948L..13J,2024ApJ...960...25K,2025MNRAS.537..402K}.
In parallel, the Atacama Large Millimeter/Submillimeter Array (ALMA) enables high-resolution observations of the cold interstellar medium (ISM), including dust and molecular gas.  Together, these facilities allow us to connect stellar growth to where and how efficiently gas is consumed, as well as how dust forms and evolves within galaxies \citep[e.g.,][]{2023ApJ...948L...8R,2024ApJ...968...15L,2024Natur.636...69T,2025ApJ...978..165H}.
However, spatially resolved investigations rely on a robust understanding of galaxy-wide properties such as stellar mass, star formation rate (SFR), molecular gas content, and metallicity.  Establishing these global scaling relations is therefore a prerequisite for interpreting resolved structures.

Over the past $\sim 10$ Gyrs, from the peak epoch of cosmic SFR density ($z \sim 2$--3) to the present day, the majority of star formation has occurred along the star-forming main sequence (MS), a tight relationship between stellar mass ($M_\ast$) and SFR \citep[e.g.,][]{2007ApJ...670..156D,2007ApJ...660L..43N}, while starbursts have contributed only a minor fraction ($\sim 10\%$) to the total star formation budget \citep{2011ApJ...739L..40R,2012ApJ...747L..31S}.  The normalization of the MS increases with redshift, while its scatter at fixed $M_\ast$ remains nearly constant at $\sim 0.3$~dex \citep[e.g.,][]{2014ApJ...795..104W,2014ApJS..214...15S,2023MNRAS.519.1526P}.  The existence of this tight sequence suggests a self-regulated mode of star formation governed by gas accretion, depletion, and feedback processes, although the physical origin of the scatter remains to be identified \citep[e.g.,][]{2012MNRAS.421...98D,2013ApJ...772..119L,2014MNRAS.438..262P}.

Molecular gas, the direct fuel for star formation \citep[e.g.,][]{2012ARA&A..50..531K,2022ARA&A..60..319S}, plays a central role in this regulation.  Large observational programs have established scaling relations linking SFR, molecular gas mass ($\Mgas$), and depletion time ($\taudep = \Mgas/\SFR$) across cosmic time (see \citealt{2020ARA&A..58..157T} and \citealt{2020ARA&A..58..661F} for recent reviews).
In this framework, deviations from the MS are understood as arising from coupled variations in molecular gas mass ratio ($\mugas=\Mgas/M_\ast$) and star formation efficiency ($\SFE = 1/\taudep$) \citep{2012ApJ...760....6M,2014ApJ...793...19S,2015ApJ...800...20G,2015ApJ...812L..23S,2018ApJ...853..179T}.  
Specifically, galaxies above the MS tend to exhibit both enhanced gas fractions and shorter depletion times.

Despite this progress, significant uncertainties remain at $z>1$.  Many scaling relations are derived from compilations spanning wide redshift ranges and multiple gas tracers, often combining samples with different selection functions.  In addition, individual measurements remain limited in number, and stacking analyses are frequently required to reach sufficient depth \citep[e.g.,][]{2014A&A...562A..30S,2015A&A...573A.113B,2016ApJ...820...83S,2020ApJ...901...79A,2020ARA&A..58..157T}.
Individual molecular gas measurements for statistically well-defined samples at cosmic noon are still limited \citep{2013ApJ...768...74T,2019A&A...622A.105F,2021MNRAS.508.5217P}.  Moreover, the conversion from dust mass to molecular gas mass often assumes constant gas-to-dust ratios, whereas metallicity variations can introduce additional systematic uncertainties.

Recent high-resolution ALMA studies have shown that massive MS galaxies at $z\sim1$--2 often host dust continuum and/or molecular gas (CO) emission that is significantly more compact than the stellar mass distribution, indicating centrally concentrated, high-surface-density star formation \citep[e.g.,][]{2019ApJ...877L..23P,2020ApJ...901...74T,2021MNRAS.508.5217P}.  Such structural compactness may represent an additional dimension underlying the observed MS scatter, motivating robust measurements of galaxy-wide gas content and depletion time.

In this work, we revisit the gas scaling relations around the MS using ALMA Band 7 observations of a homogeneous sample of massive ($M_\ast \gtrsim 10^{10.8}~M_\odot$) star-forming galaxies at $1.4<z<1.7$.  
Our targets are drawn from the FMOS-COSMOS survey \citep{2015ApJS..220...12S,2019ApJS..241...10K}, which provides secure spectroscopic redshifts, extinction-corrected H$\alpha$-based SFRs, and individual gas-phase metallicity measurements.  The narrow redshift interval minimizes evolutionary mixing, enabling a controlled test of MS scaling relations at cosmic noon.

The ALMA observations are sufficiently deep to yield dust continuum detections for nearly the entire sample, allowing molecular gas masses to be inferred individually.  Crucially, we adopt metallicity-dependent gas-to-dust ratios for individual galaxies, reducing systematic uncertainties associated with gas mass estimates.  This combination of deep individual detections, homogeneous selection, and metallicity-informed gas conversion provides a clean probe of how molecular gas fraction and depletion time govern the scatter of the MS in massive galaxies.

The structure of the paper is as follows.  Section \ref{sec:sample_data} describes the sample selection and observations.
Section \ref{sec:derivation} presents the derivation of galaxy and ISM properties.  Results and discussion are given in Section \ref{sec:results}.  
We summarize in Section \ref{sec:summary}.

Throughout the paper we adopt a flat $\Lambda$CDM cosmology with $\Omega_\mathrm{\Lambda}=0.69$, $\Omega_\mathrm{M}=0.31$ and  $H_0=67.7~\mathrm{km~s^{-1}~Mpc^{-1}}$ \citep{2020A&A...641A...6P}.  Magnitudes are quoted in the AB system and a \citet{2003PASP..115..763C} initial mass function (IMF) is assumed.

\section{Sample and data}\label{sec:sample_data}

This work is based on a sample of massive star-forming galaxies selected from the FMOS-COSMOS survey \citep{2015ApJS..220...12S,2019ApJS..241...10K}.  Below, we briefly summarize the parent survey, describe the selection of the subsample analyzed in this work, and present the ALMA observations and dust continuum flux measurements.

\subsection{The FMOS-COSMOS survey}
\label{sec:FMOS-COSMOS}

The FMOS-COSMOS survey is a large spectroscopic campaign in the COSMOS field \citep{2007ApJS..172....1S}, conducted with the Fiber Multi-Object Spectrograph (FMOS; \citealt{2010PASJ...62.1135K}) on the Subaru Telescope \citep{2015ApJS..220...12S,2019ApJS..241...10K}.  
The primary observations were carried out with the $H$-long grating ($\lambda = 1.6$--$1.8~\mathrm{\mu m}$), targeting H$\alpha$ and [\Nii]$\lambda$6584 emission lines in star-forming galaxies at $1.43 < z < 1.73$.  A subset ($\sim 35$\%) of targets observed in $H$-long were further observed with the $J$-long grating ($1.11\textrm{--}1.35~\mathrm{\mu m}$) to measure H$\beta$+[\Oiii]$\lambda$4959,5007 emission lines.

The parent FMOS-COSMOS sample was constructed mainly from $K$-selected galaxies with photometric redshifts appropriate for the FMOS spectral coverage.  Expected H$\alpha$ fluxes, inferred from SED-based estimates of SFR, were also used in the target selection.  In addition, FIR-bright galaxies identified in the Herschel/PACS Evolutionary Probe (PEP) survey \citep{2011A&A...532A..90L} were included to extend the sampling of starburst systems (see also \citealt{2015ApJ...806L..35K}).
While the inclusion of these FIR-bright sources mainly added galaxies at relatively lower stellar masses ($M_\ast < 10^{10.5}~M_\odot$), it had little impact in the mass range relevant to this study, $M_\ast \gtrsim 10^{10.8}~M_\odot$, because massive FIR-bright galaxies largely overlap with normal main-sequence star-forming galaxies. Therefore, in this mass range, our selection based on stellar mass and SFR is expected to provide a representative sample of massive main-sequence galaxies.

\subsection{Sample selection for the ALMA Band 7 observations}
\label{sec:selection}

From the FMOS-COSMOS survey, we select massive star-forming galaxies for ALMA Band 7 observations based on the following criteria: i) secure spectroscopic redshift from FMOS-COSMOS based on H$\alpha$(+[\Nii]) detection ($\mathrm{zflag}\ge3$), ii) dust-corrected H$\alpha$-based SFR $\ge 20 M_\odot~\mathrm{yr^{-1}}$, iii) stellar mass $M_\ast \gtrsim 10^{10.8}~M_\odot$, and iv) lying within the central 0.6~deg$^2$ area expected at the time to be covered by the COSMOS-Web Large Program \citep{2023ApJ...954...31C}.\footnote{At the time of sample selection, however, the COSMOS-Web footprint had not yet been finalized, and seven sources are not covered by the final NIRCam footprint.}  Sources detected in X-rays in the \textit{Chandra} COSMOS Legacy Survey \citep{2016ApJ...819...62C} are excluded to minimize AGN contamination.  The resulting sample contains 57 galaxies.

Among the 57 FMOS-ALMA targets, 20 are identified in the PEP catalog.  In our SED fitting analysis, we adopt FIR photometry from \citet{2018ApJ...864...56J} rather than the original PEP catalog (see Section \ref{sec:sed-fitting}).  With these measurements, 32 sources are detected at $>3\sigma$ in at least one of the PACS (100, 160~$\mu m$) or SPIRE (250, 350, 500~$\mu m$) bands.
Thus, the sample is not limited to FIR-bright galaxies, and the new ALMA Band 7 continuum measurements provide essential constraints on the dust emission for the majority of the sample.

Two sources (INDEX 1871 and 3550) exhibit broad H$\alpha$ emission ($>1000~\mathrm{km~s^{-1}}$), indicative of possible AGN activity. Although they were observed with ALMA, we exclude them from the analysis presented here in order to focus the star-forming galaxy sample. Another source (INDEX 2076) is also excluded from the analysis because the SED fitting does not converge to a reasonable solution (see Section \ref{sec:sed-fitting}).\footnote{The photometric fluxes in the rest-frame optical-to-NIR window are of poor quality, and the best-fit SED yields an extremely high reduced $\chi^2$ of $\approx 12$.} 
The final statistical sample therefore consists of 54 galaxies.

\begin{figure}
 \begin{center}
 \includegraphics[width=3.4in]{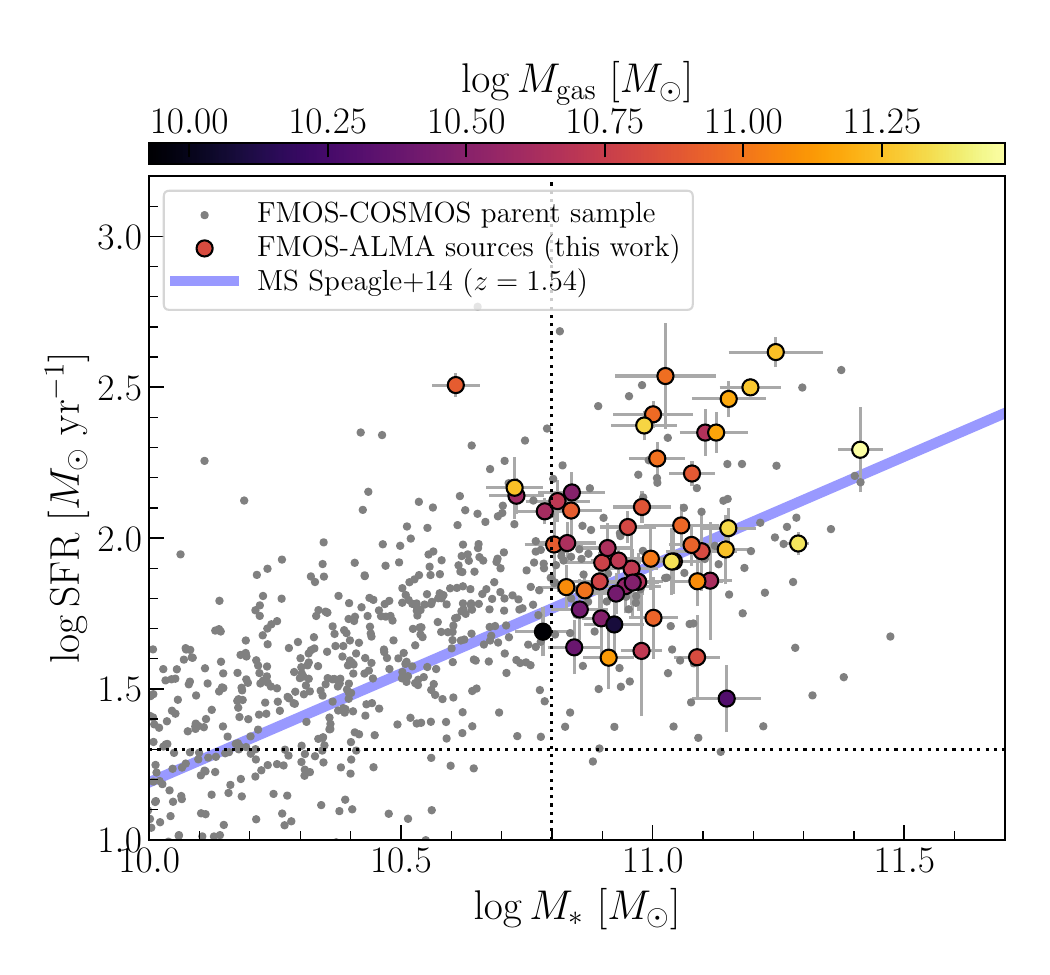}
 \end{center}
\caption{Stellar mass versus SFR for the FMOS-ALMA sample, with both derived from the CIGALE SED fitting described in Section~\ref{sec:sed-fitting}.  The gray distribution shows all FMOS-COSMOS galaxies at $1.4<z<1.7$ with secure H$\alpha$ detection.  The solid line indicates the star-forming main sequence evaluated at the median redshift of the ALMA sample ($z = 1.53$) following \citet{2014ApJS..214...15S}.  The ALMA targets occupy the massive end of the main sequence with no significant bias in SFR across the MS.
}\label{fig:Mstar_vs_SFR}
\end{figure}

Figure \ref{fig:Mstar_vs_SFR} shows the distribution of these galaxies in the  $M_\ast$--SFR plane relative to the main sequence at this epoch \citep{2014ApJS..214...15S}, together with the FMOS-COSMOS parent sample.  The ALMA sample occupies the massive end of the star-forming MS without significant bias in SFR, sampling both the upper and lower sides of the sequence.

We note that the stellar masses and SFRs have been updated from the values used at the time of ALMA target selection, based on revised SED fitting that incorporates the latest photometric data and ALMA flux measurements (see Section~\ref{sec:sed-fitting}). As a consequence, a small number of sources now fall below the original stellar mass threshold of $10^{10.8}~M_\odot$ adopted for the target selection.

\subsection{ALMA observations and continuum flux measurements}

Of the 57 selected targets, 56 were observed in an ALMA Cycle 8 program (2021.1.01133.S; PI D. Kashino).  The observations were defined to detect dust continuum emission at 334~GHz ($897~\mu$m), corresponding to rest-frame $\sim 350~\mathrm{\mu}$m at $z\sim1.5$, in Band 7.  Four spectral windows of 1.875~GHz were used, providing a total bandwidth of 7.5~GHz.  

The targets were observed between 24 and 29 September 2022 in the C-3 configuration using 42 12-m array antennas.  The on-source integration times range from 209 to 314 seconds.  The bandpass and phase calibrators were J0948$+$0022 and J1010$-$0200.

The data were processed using the standard ALMA CASA pipeline (version 6.4.1.12).  Continuum images were produced with the \texttt{tclean} task using Briggs weighting with a robust parameter of 0.5.  The resulting synthesized beam size is approximately $0.56\times0.44$~arcsec ($\approx 4.9\times3.8~\mathrm{kpc}$ at $z=1.5$) and the achieved rms is $0.073~\mathrm{mJy/beam}$ over the 7.5~GHz bandwidth.

One target (INDEX 612) is covered by archived Band 7 observations originally targeting a nearby source (DSFGS2.95; Project code 2015.1.00568.S; PI C. Casey).  We retrieved the corresponding ARI-L\footnote{The Additional Representative Images for Legacy (ARI-L; \url{https://almascience.nrao.edu/alma-data/aril}).} cleaned continuum image at a representative frequency of 344~GHz.  The synthesized beam size is comparable to the main dataset, with an rms sensitivity of 0.12~mJy/beam over 7.5~GHz.

Dust continuum flux densities for the main ALMA Cycle 8 dataset are measured via visibility-plane ($uv$-plane) fitting (see \citealt{2024A&A...684A..23T,2025MNRAS.536.3090K}), which we adopt as the fiducial measurements.  For consistency checks, we also perform aperture photometry on the cleaned images using circular apertures of 3~arcsec diameter, and confirm that the results agree within the uncertainties.  For the archival source (INDEX 612), we measure the flux density from the cleaned continuum image using a 3~arcsec diameter circular aperture centered on the target position. Given the beam size and sensitivity comparable to the main dataset, this measurement is directly consistent with the fluxes derived for the primary sample.  The typical uncertainty of the Band 7 flux measurements is 0.14~mJy.  Dust continuum emission is detected at $\ge3\sigma$ ($\ge2\sigma$) significance for 55 (all 57) sources.  Table~\ref{tab:catalog1} summarizes the FMOS-ALMA sample and the Band 7 continuum flux measurements.

\begin{table}
  \tbl{ALMA Band 7 continuum flux measurements$^*$}{%
  \begin{tabular}{ccccc}
      \hline
      INDEX &
      R.A. &
      Decl. &
      Redshift &
      $f_{\mathrm{Band7}}$$^\dag$ \\
      &
      (degree) &
      (degree) &
      &
      (mJy) \\
      \hline
81 & 150.164430 & 1.936306 & 1.525 & $0.95 \pm 0.14$ \\
89 & 149.855774 & 2.130164 & 1.478 & $0.37 \pm 0.12$ \\
127 & 149.771467 & 1.915945 & 1.534 & $0.50 \pm 0.10$ \\
147 & 149.745818 & 2.125955 & 1.556 & $1.28 \pm 0.11$ \\
193 & 149.925367 & 2.135018 & 1.470 & $0.91 \pm 0.14$ \\
275 & 149.908679 & 2.482474 & 1.567 & $1.17 \pm 0.19$ \\
281 & 150.124035 & 2.447523 & 1.599 & $1.03 \pm 0.14$ \\
285 & 149.972726 & 2.490150 & 1.455 & $0.88 \pm 0.12$ \\
316 & 149.981431 & 2.253202 & 1.445 & $2.58 \pm 0.09$ \\
326 & 149.912289 & 2.281628 & 1.550 & $1.17 \pm 0.12$ \\
      \hline
  \end{tabular}}
  \label{tab:catalog1}
\begin{tabnote}
$^*$ The full version of this table will be available in machine-readable form. A representative portion is shown here to illustrate its format and content. \\
$^\dag$ ALMA Band 7 continuum flux density.
\end{tabnote}
\end{table}

\begin{figure*}[t]
\begin{center}
\includegraphics[width=3.4in]{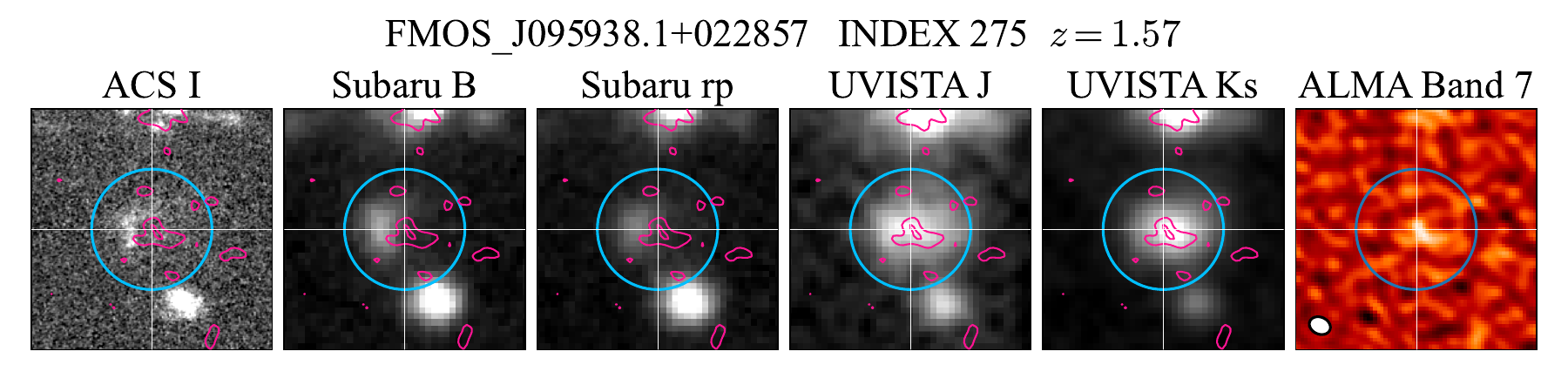}
\includegraphics[width=3.4in]{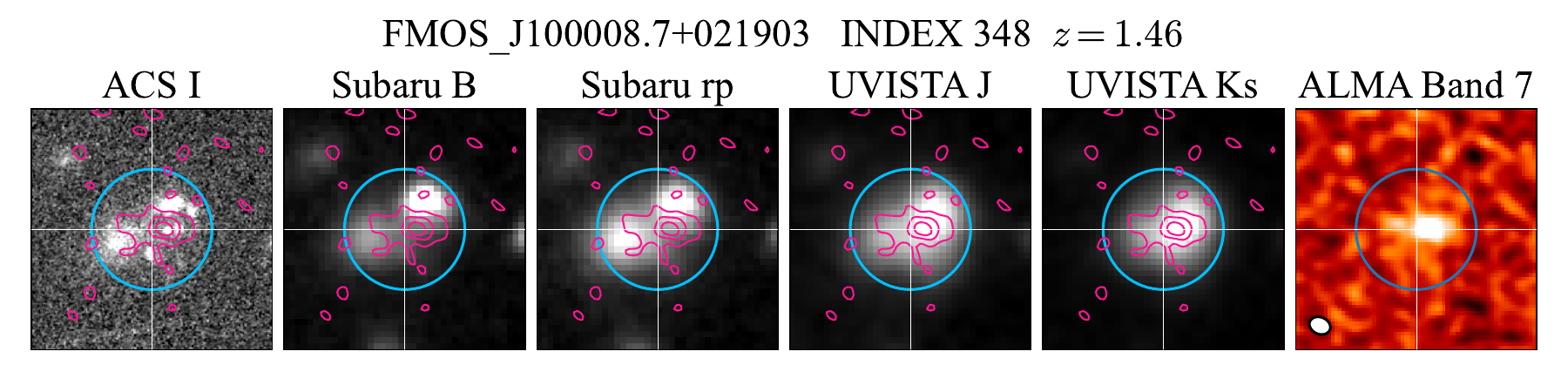}
\includegraphics[width=3.4in]{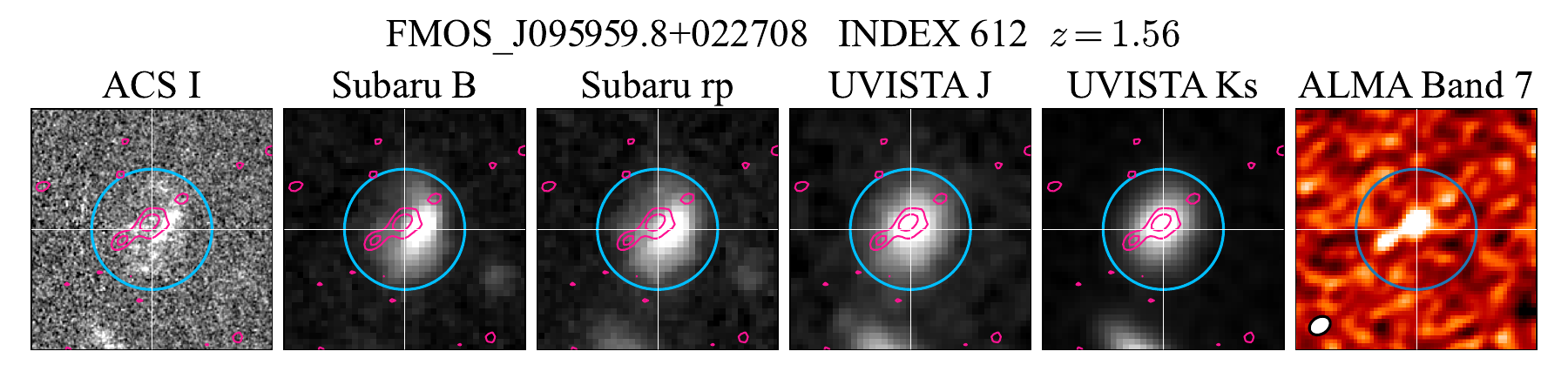}
\includegraphics[width=3.4in]{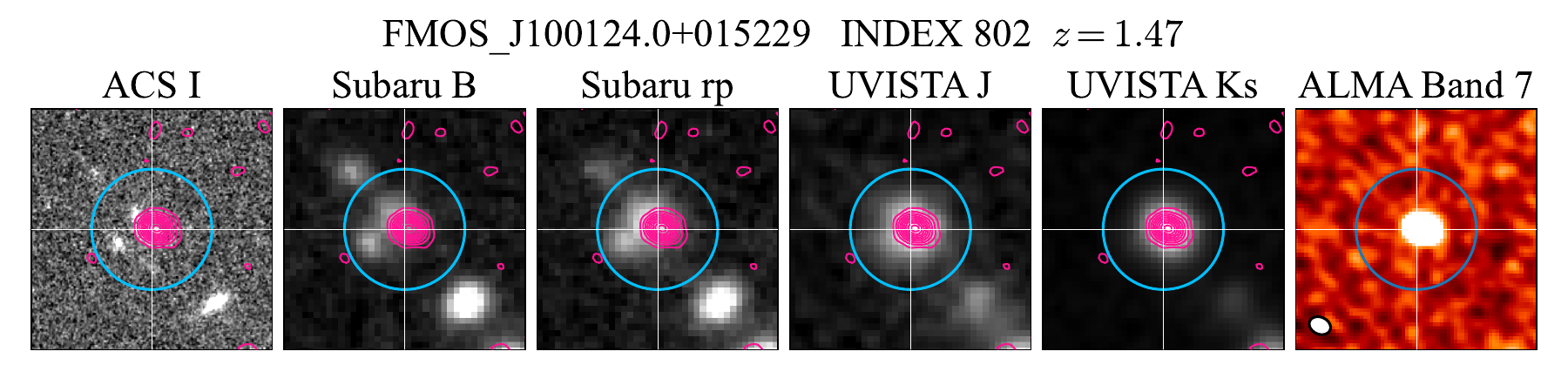}
\includegraphics[width=3.4in]{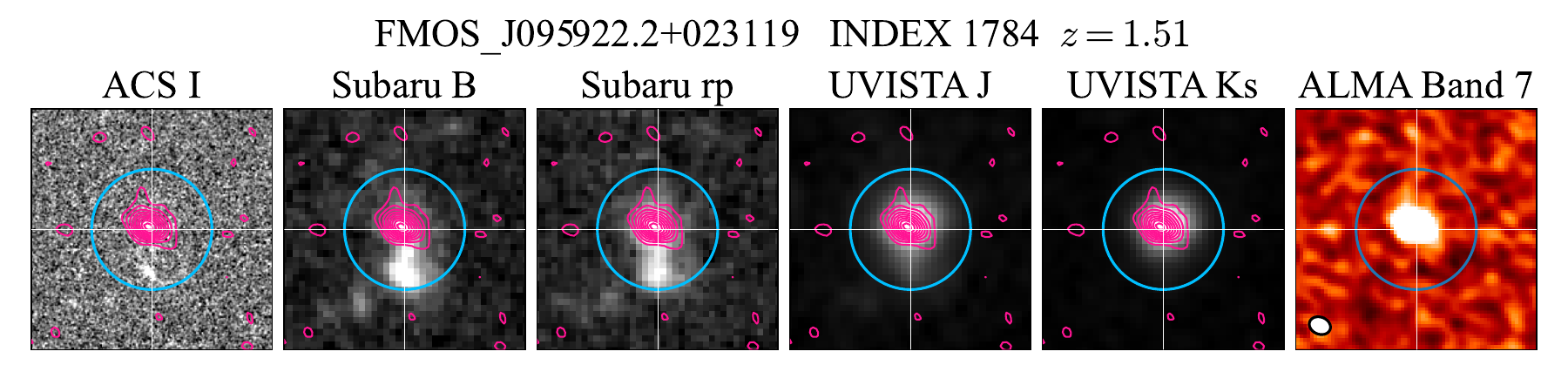}
\includegraphics[width=3.4in]{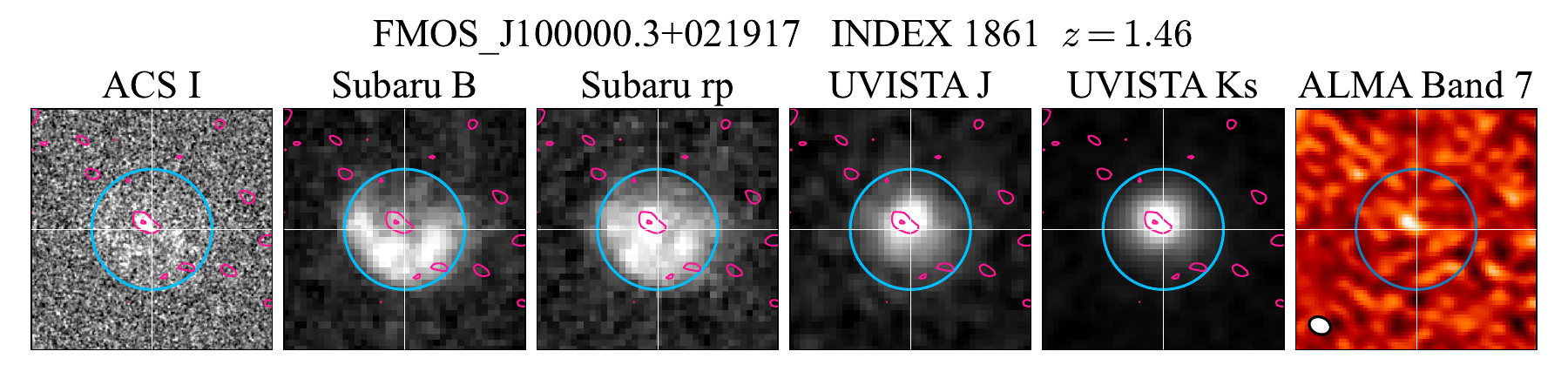}
\end{center}
\caption{Example cutout images of representative FMOS-ALMA sources (6\arcsec on a side; north is up and east is left).  From left to right, the HST/ACS F814W, Subaru/Suprime-Cam $B$ and $r+$, VISTA/VIRCam $J$ and $K_\mathrm{s}$, and ALMA Band 7 continuum images are shown.  The synthesized ALMA beam is indicated in the lower-left corner of the rightmost panels. Contours overlaid on the optical-to-near-IR images trace the Band 7 dust continuum emission at levels of $1\sigma$, $3\sigma$, $5\sigma$, and so on.  The 3\arcsec-diameter aperture is shown as a cyan circle in each panel.
\label{fig:stamps}}
\end{figure*}

Figure~\ref{fig:stamps} presents example cutout images of the FMOS-ALMA sources, comparing the ALMA dust continuum maps with optical and near-infrared imaging from HST/F814W \citep{2007ApJS..172..196K}, Subaru/Suprime-Cam, and VISTA/VIRCam \citep{2012A&A...544A.156M}, retrieved from the COSMOS Archive.\footnote{\url{https://irsa.ipac.caltech.edu/data/COSMOS/}}
Although the present study uses only integrated flux measurements, these examples illustrate the diversity of morphologies and the varying spatial correspondence between dust emission and stellar light at shorter wavelengths.  A detailed analysis of the resolved stellar and ISM structures will be presented in companion papers (\citealt{2025MNRAS.536.3090K}; Suzuki et al., in prep).

\section{Derivation of galaxy and ISM properties} \label{sec:derivation}

In this section, we describe how the key galaxy and ISM properties are derived for the FMOS-ALMA sample. 

\subsection{SED Fitting and Derivation of Galaxy Properties}
\label{sec:sed-fitting}

\begin{table*}
  \tbl{Inferred properties of the FMOS-ALMA sources$^*$}{%
  \small
  \begin{tabular}{cccccccc}
      \hline
      INDEX &
      $\log M_\ast$ &
      $\log \mathrm{SFR}$ &
      $12+\log \mathrm{(O/H)}$ &
      $\log M_{\mathrm{dust}}$ &
      $\log L_{\mathrm{IR}}$ &
      $\langle U\rangle$ &
      $\log M_{\mathrm{gas}}$ \\
      &
      ($M_\odot$) &
      ($M_\odot\,\mathrm{yr}^{-1}$) &
      (PP04)&
      ($M_\odot$) &
      ($L_\odot$) &
      &
      ($M_\odot$) \\
      \hline
81 & $10.91 \pm 0.05$ & $1.60 \pm 0.10$ & $8.78 \pm 0.06$ & $9.20 \pm 0.16$ & $11.65 \pm 0.07$ & $2.6 \pm 1.6$ & $11.13 \pm 0.17$ \\
89 & $11.15 \pm 0.07$ & $1.47 \pm 0.11$ & $8.72 \pm 0.10$ & $8.31 \pm 0.29$ & $11.58 \pm 0.11$ & $20.6 \pm 19.5$ & $10.30 \pm 0.31$ \\
127 & $10.93 \pm 0.05$ & $1.82 \pm 0.07$ & $8.60 \pm 0.07$ & $8.33 \pm 0.11$ & $11.82 \pm 0.08$ & $23.7 \pm 6.3$ & $10.42 \pm 0.13$ \\
147 & $11.08 \pm 0.05$ & $1.98 \pm 0.07$ & $8.58 \pm 0.07$ & $8.82 \pm 0.12$ & $12.08 \pm 0.04$ & $14.3 \pm 4.2$ & $10.93 \pm 0.13$ \\
193 & $10.95 \pm 0.06$ & $2.04 \pm 0.05$ & $8.57 \pm 0.04$ & $8.69 \pm 0.15$ & $12.03 \pm 0.03$ & $18.2 \pm 6.5$ & $10.80 \pm 0.15$ \\
275 & $10.81 \pm 0.06$ & $2.12 \pm 0.07$ & $8.72 \pm 0.08$ & $8.72 \pm 0.13$ & $12.14 \pm 0.05$ & $20.8 \pm 5.8$ & $10.71 \pm 0.15$ \\
281 & $11.10 \pm 0.07$ & $1.96 \pm 0.13$ & $8.73 \pm 0.03$ & $8.84 \pm 0.18$ & $11.99 \pm 0.16$ & $13.4 \pm 10.0$ & $10.82 \pm 0.18$ \\
285 & $10.97 \pm 0.05$ & $1.86 \pm 0.10$ & $8.87 \pm 0.06$ & $8.80 \pm 0.16$ & $11.94 \pm 0.06$ & $12.1 \pm 6.0$ & $10.66 \pm 0.17$ \\
316 & $10.61 \pm 0.05$ & $2.51 \pm 0.04$ & $8.81 \pm 0.02$ & $8.98 \pm 0.03$ & $12.44 \pm 0.03$ & $20.9 \pm 2.5$ & $10.90 \pm 0.03$ \\
326 & $10.98 \pm 0.06$ & $2.10 \pm 0.05$ & $8.68 \pm 0.03$ & $8.85 \pm 0.13$ & $12.12 \pm 0.04$ & $15.0 \pm 5.1$ & $10.87 \pm 0.14$ \\      
    \hline
  \end{tabular}}
  \label{tab:catalog2}
\begin{tabnote}
$^*$ This table is published in its entirety in machine-readable format. A portion is shown here to illustrate its format and content.
\end{tabnote}
\end{table*}

To derive the physical properties of each galaxy, we perform spectral energy distribution (SED) fitting using multiwavelength photometric data together with the ALMA dust continuum measurements.

We adopt photometric fluxes from the GALEX near-UV to Spitzer/IRAC channel 4 from the COSMOS2020 catalog \citep{2022ApJS..258...11W}.  For CFHT, Subaru, and VISTA photometry, we use fluxes measured within 3\arcsec-diameter apertures and apply Galactic extinction corrections following \citet{2022ApJS..258...11W}.  
We incorporate the super-deblended FIR photometric catalog provided by \citet{2018ApJ...864...56J} for Spitzer/MIPS 24~$\mu$m and Herschel/PACS (100, 160~$\mu$m) and SPIRE (250, 350, 500~$\mu$m) bands.  The ALMA Band 7 continuum flux at $\lambda_{\rm obs}=897~\mu$m is included at the longest wavelength.

Flux densities and their associated uncertainties are used directly in the fitting procedure, rather than magnitudes.  This approach allows non-detected measurements to be incorporated self-consistently within the likelihood framework without requiring a separate treatment of upper limits.

We perform SED fitting using CIGALE (v2022.1; \citealt{2019A&A...622A.103B}), adopting the stellar population synthesis models of \citet{2003MNRAS.344.1000B} with a \citet{2003PASP..115..763C} IMF and a fixed stellar metallicity of $Z_\ast=0.008$.  Delayed star formation histories with an additional recent burst of 20~Myr are assumed to describe both long-term stellar mass assembly and recent star formation activity.

Nebular emission is included assuming ionization parameter\footnote{This ionization parameter, $U_\mathrm{ion}$, should not be confused with the radiation field parameters $U_{\min}$ and $U_{\max}$ that appear later in the dust emission model (Section \ref{sec:derive_Mdust}). These quantities are unrelated to each other, and $U_\mathrm{ion}$ is used only to determine the nebular emission.}  $\log U_{\rm ion} = -2.8$, gas metallicity $Z_\mathrm{gas}=0.008$, and electron density $100~\mathrm{cm^{-3}}$. The escape fraction of Lyman continuum photons and dust absorption within H\,\textsc{ii} regions are set to zero.

Dust attenuation follows the modified starburst prescription implemented in CIGALE, based on the \citet{2000ApJ...533..682C} curve and extended with \citet{2002ApJS..140..303L} shortward of 1500\,\AA. The overall slope can vary with a power-law modification with a variable index, and a UV bump of variable amplitude can be included. 

The full set of variable parameters and their adopted ranges in the CIGALE fitting are summarized in Table~\ref{tab:cigale} in Appendix \ref{sec:apx_cigale}. 
These include parameters governing the star formation history, dust attenuation, nebular emission, and dust emission components (see Section \ref{sec:derive_Mdust}). 
The chosen parameter ranges are sufficiently broad to explore a wide range of physically plausible solutions.

For all sources retained in the final analysis sample, the SED fits converge successfully and yield well-constrained galaxy properties.  We adopt Bayesian estimates provided by CIGALE, together with their associated uncertainties, for all subsequent analyses.  The SFRs are averaged over 10~Myr (\texttt{sfr10Myrs}) to trace recent star formation in a manner consistent with the burst component allowed in our SFH model.  Table~\ref{tab:catalog2} summarizes the inferred physical properties.

The derived stellar masses span $10^{10.7}$--$10^{11.4}~M_\odot$, and the SFRs range from $10^{1.4}$--$10^{2.5}~M_\odot~\mathrm{yr^{-1}}$.  Figure~\ref{fig:Mstar_vs_SFR} shows the derived stellar masses and SFRs, compared with the parent FMOS-COSMOS sample.  For the parent sample, stellar masses and SFRs are estimated from SED fitting using COSMOS2015 photometric data with a simpler dust emission model \citep{2014ApJ...784...83D}.  The parent sample is shown for reference only and is not used in the present analysis.

For all subsequent analyses, we adopt the SFRs derived from the SED fitting.  As discussed in Appendix \ref{sec:apx_cigale}, these SED-based SFRs are broadly consistent with those estimated from dust-corrected H$\alpha$ luminosities in the FMOS-COSMOS survey \citep{2019ApJS..241...10K} once aperture correction is applied, although substantial scatter remains.  This scatter likely reflects uncertainties in dust attenuation and aperture corrections to the H$\alpha$ luminosities, with the latter contributing a typical uncertainty of 0.17~dex \citep{2019ApJS..241...10K}.  In addition, each individual H$\alpha$ measurement depends on the pointing accuracy of the corresponding fiber, and it is not possible to correct rigorously for this effect on a source-by-source basis.  Therefore, the true uncertainties in the H$\alpha$-based SFRs may be substantially larger than the typical uncertainties in the SED-based SFRs.  Given these considerations, we adopt SED-based SFRs for all subsequent analyses.

While stellar mass and SFR are primarily constrained by the UV–optical part of the SED, the far-infrared to submillimeter regime provides direct constraints on the dust mass. We describe the derivation of $\Mdust$ in the next subsection.

\subsection{Derivation of the Dust Mass}
\label{sec:derive_Mdust}

\begin{figure*}[p]
\begin{center}
\includegraphics[width=3.0in]{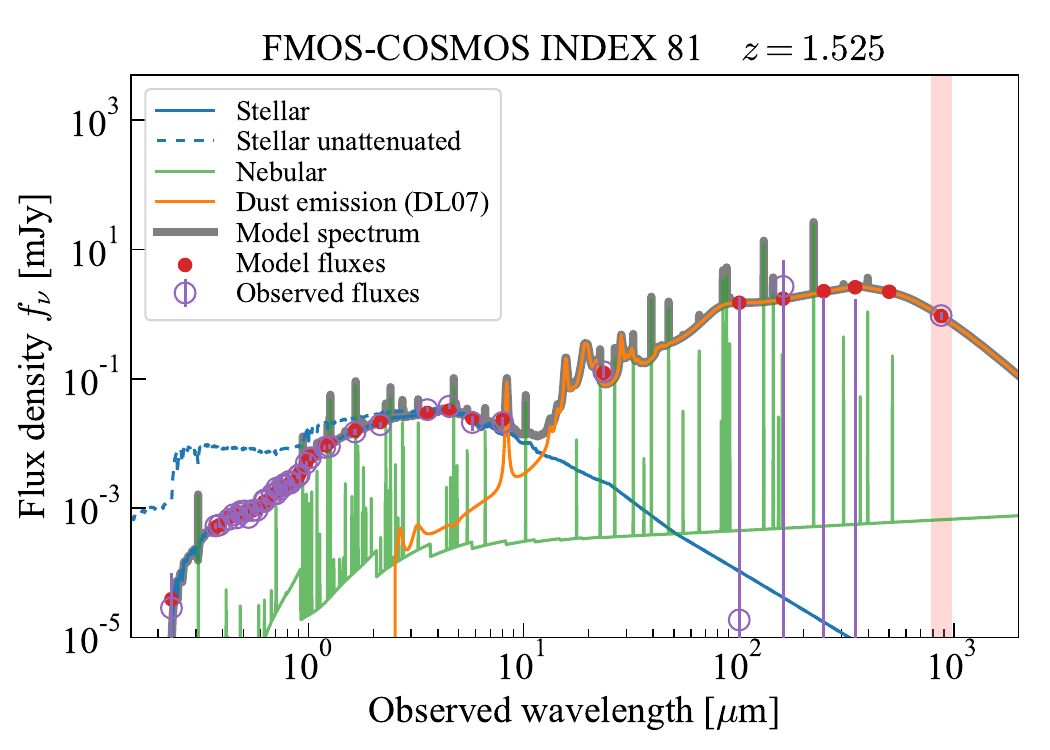} 
\includegraphics[width=3.0in]{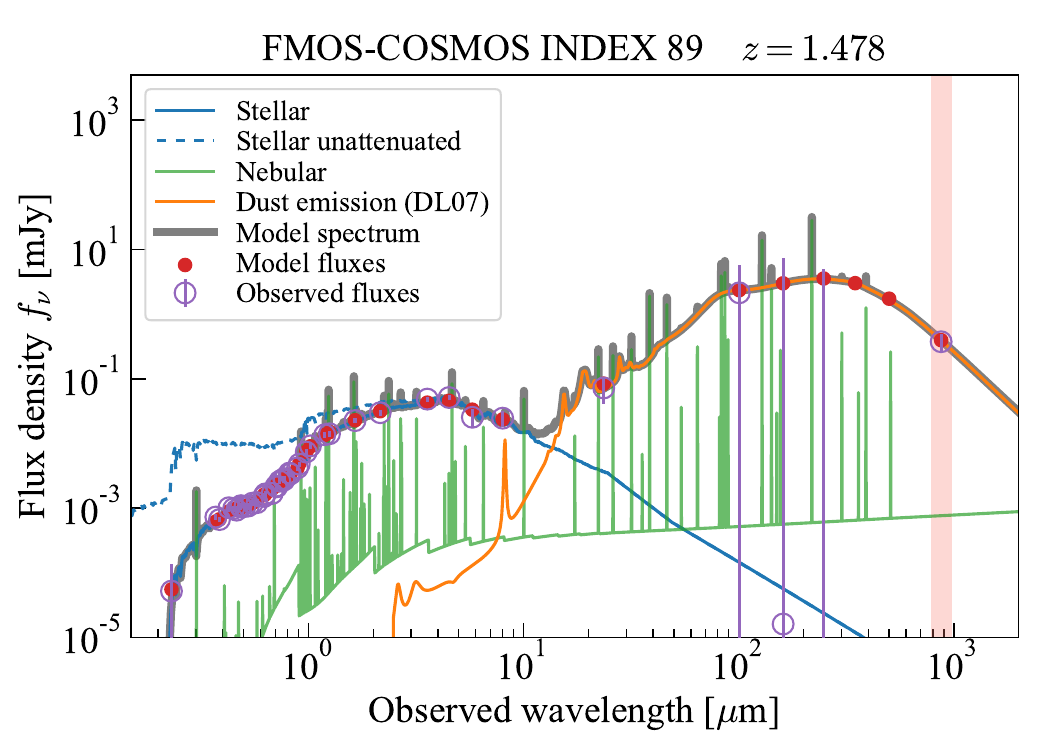} 
\includegraphics[width=3.0in]{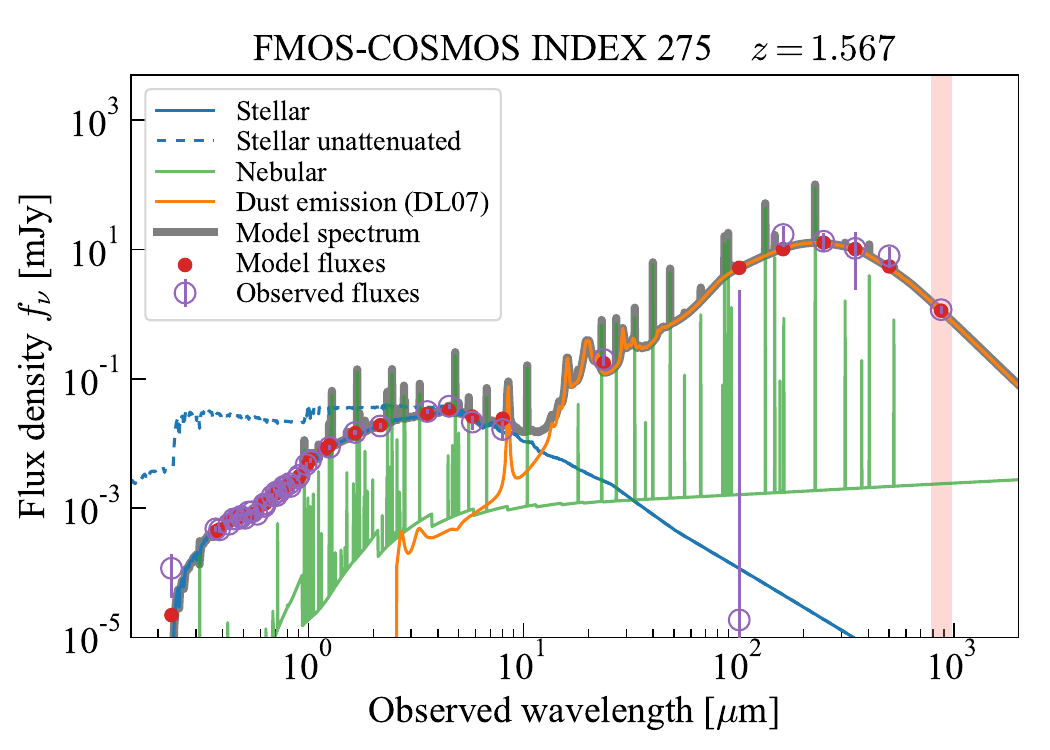} 
\includegraphics[width=3.0in]{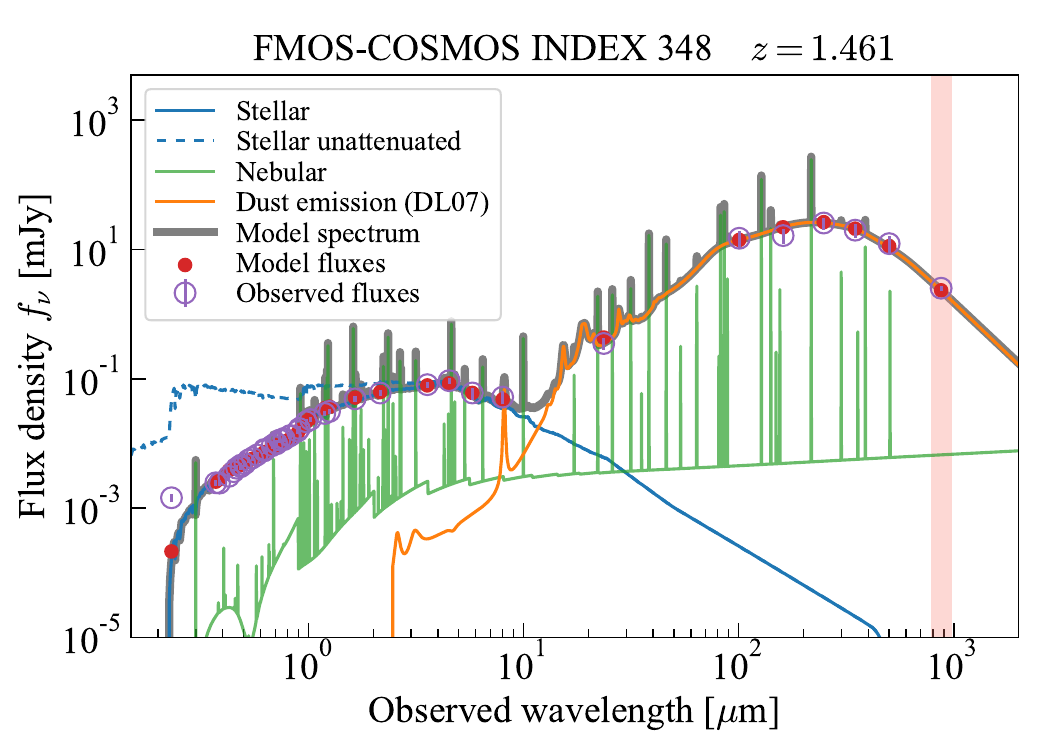} 
\includegraphics[width=3.0in]{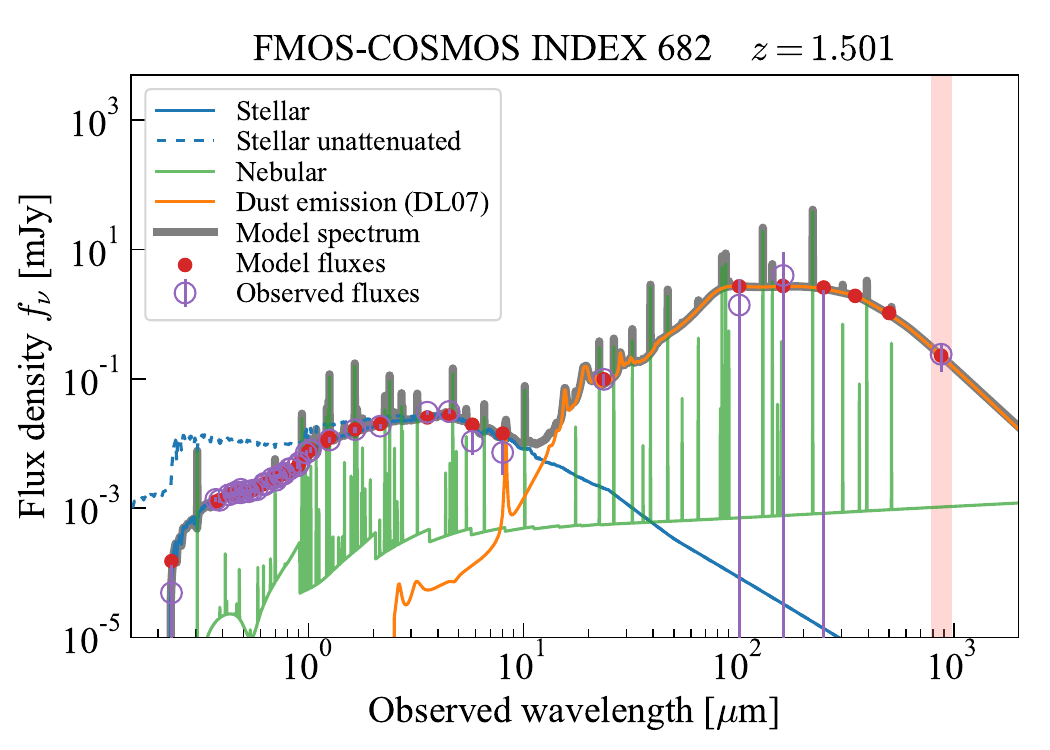} 
\includegraphics[width=3.0in]{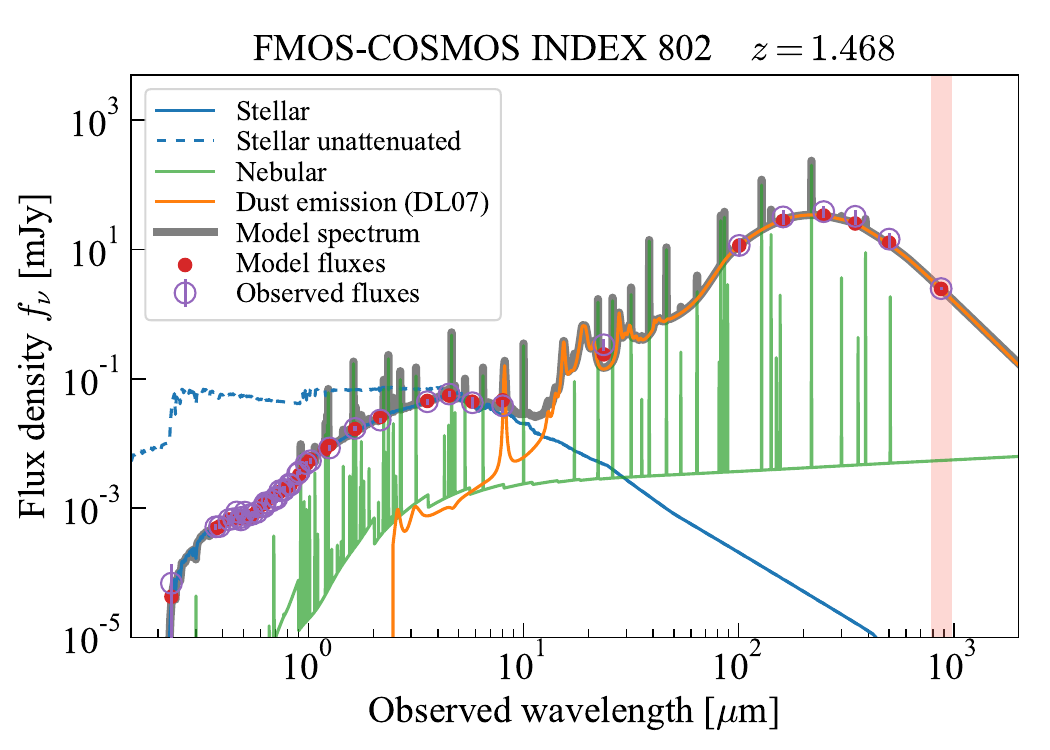} 
\includegraphics[width=3.0in]{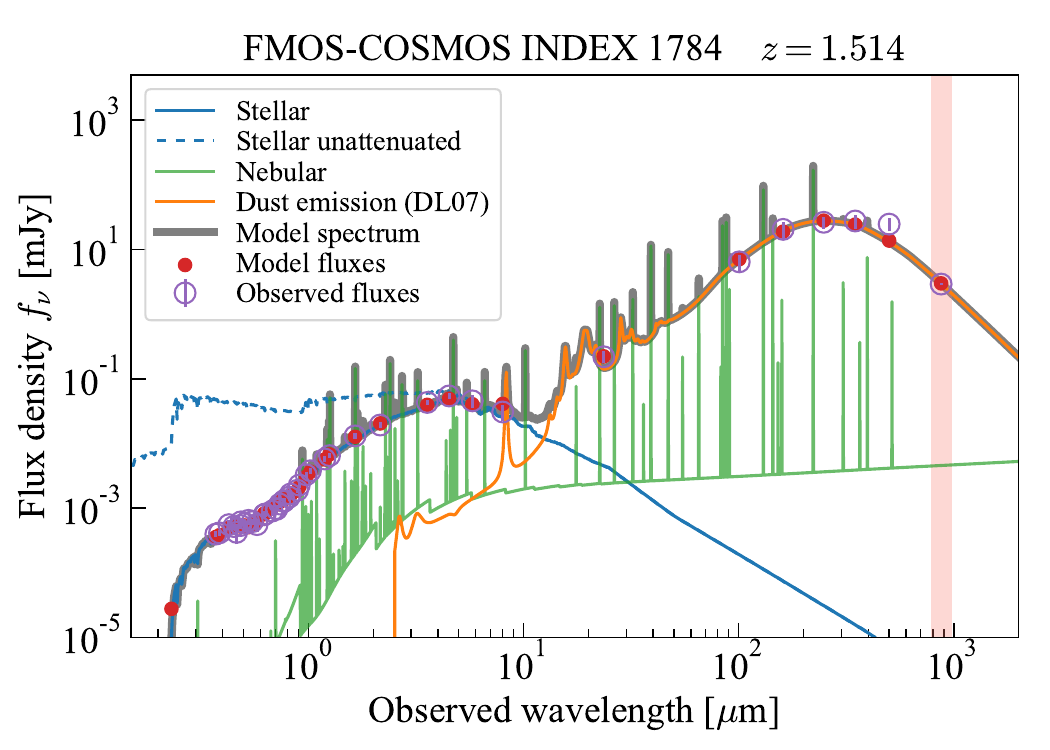} 
\includegraphics[width=3.0in]{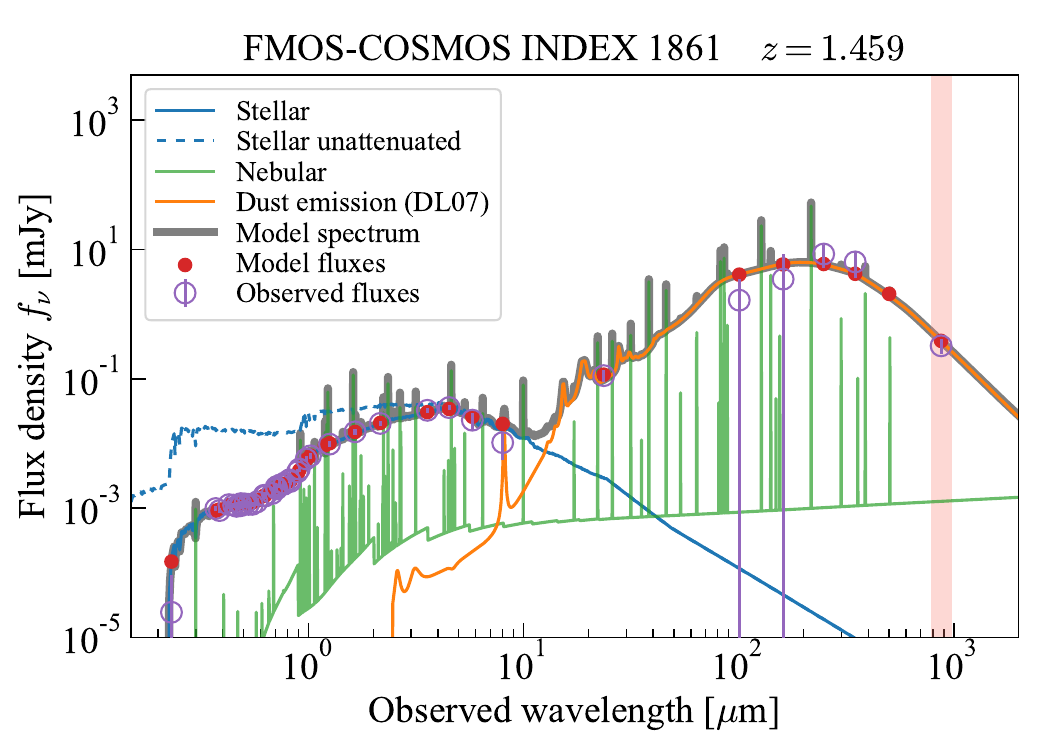} 
\end{center}
\caption{Examples of CIGALE SED fits for eight representative FMOS-ALMA galaxies. In each panel, the best-fit SED model is shown by the thick gray solid line, with its subcomponents indicated in the legend. Observed flux densities with uncertainties are shown as open circles, and model-predicted fluxes as red filled circles. The ALMA Band 7 data point is highlighted by a thick vertical line, illustrating its role in constraining the long-wavelength dust emission. This demonstrates that, with the inclusion of the ALMA data, the dust emission can be reliably constrained even for Herschel-undetected sources (INDEX 81, 89, and 682).
\label{fig:sed-fits}}
\end{figure*}

The dust mass $\Mdust$ is primarily constrained by the FIR to submillimeter part of the SED, where thermal emission from large dust grains dominates.  In particular, the inclusion of the ALMA Band 7 continuum measurement at $\lambda_{\rm obs}=897~\mu$m (rest-frame $\sim350~\mu$m at $z\sim1.5$) provides a critical constraint on the Rayleigh-Jeans tail of the dust emission spectrum.  This long-wavelength data point significantly reduces degeneracies between dust temperature and normalization, thereby improving the robustness of the infrared dust masses.

We adopt the dust emission model of \citet{2007ApJ...657..810D} as implemented in CIGALE.  In this framework, dust is heated by a distribution of interstellar radiation fields characterized by a minimum radiation field $U_\mathrm{min}$ and a power-law distribution extending to higher intensities with a fraction $\gamma$ of dust exposed to intense radiation.  The power-law index is assumed to be $-2$ and the maximum intensity $U_\mathrm{max}$ is fixed to $10^6$ in our CIGALE fitting.  The model also includes variations in the mass fraction of polycyclic aromatic hydrocarbons (PAHs), $q_\mathrm{PAH}$.  The free parameters governing the dust emission component and their adopted ranges are listed in Table~\ref{tab:cigale}.

For our sample, the combination of Herschel FIR photometry and ALMA Band 7 continuum measurements provides strong leverage on both the peak and the Rayleigh-Jeans slope of the dust SED.  Figure~\ref{fig:sed-fits} illustrates representative fits, highlighting the importance of the ALMA data point in anchoring the long-wavelength side of the spectrum.  
Including the ALMA measurement significantly improves the constraints on $M_\mathrm{dust}$, reducing its uncertainty by a factor of $\sim 10$ on average and weakening the degeneracy with $U_\mathrm{min}$.

We adopt the Bayesian estimates of $\Mdust$ from the CIGALE likelihood distributions.  The resulting dust masses span $\Mdust \sim 10^{8.0}$--$10^{9.5}~M_\odot$, with typical uncertainties of $\sim0.1$--$0.16$~dex.  The median dust mass of the sample is $\approx 10^{8.8}~M_\odot$.

The robust determination of $\Mdust$ for nearly the entire sample, enabled by the high detection rate in ALMA Band 7 (55 and 42 out of 57 sources at $\ge 3\sigma$ and $\ge 5\sigma$, respectively), sets the foundation for the derivation of molecular gas masses presented in the next subsection.

\subsection{Derivation of Molecular Gas Masses}
\label{sec:derive_Mgas}

\begin{figure}[t]
\begin{center}

\includegraphics[width=3.4in]{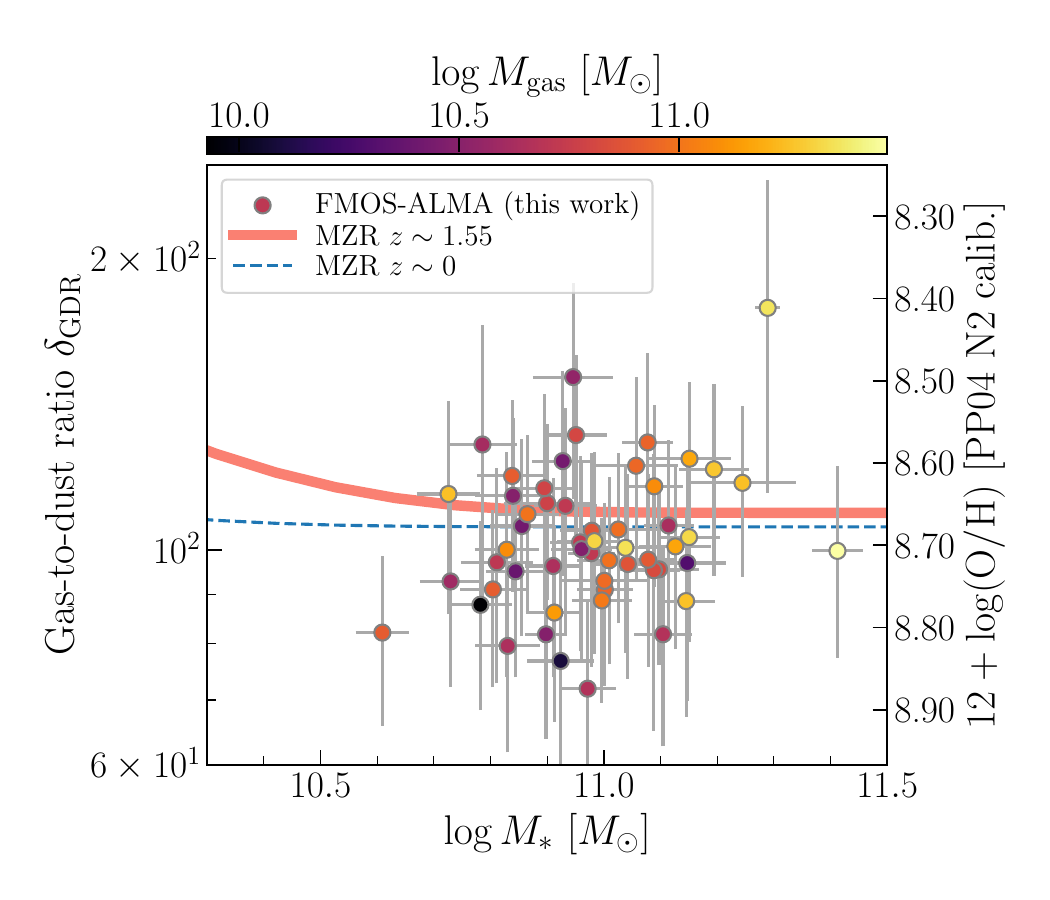}
\end{center}
\caption{Gas-to-dust ratio $\gdr$ as a function of stellar mass. Symbols are color-coded by molecular gas mass.  For reference, the corresponding metallicity is indicated on the right-hand $y$ axis, and the mass-metallicity relations (MZRs) at $z\sim0$ and $z\sim1.55$ \citep{2019ApJS..241...10K} are shown.  The inferred $\gdr$ shows little dependence on $M_\ast$ or $\Mgas$ (and equivalently $\Mdust$), reflecting the relatively flat mass-metallicity relation at the massive end.
\label{fig:Mstar_vs_GDR}}
\end{figure}

With robust dust mass estimates in hand (Section~\ref{sec:derive_Mdust}), we derive molecular gas masses by adopting a metallicity-dependent gas-to-dust mass ratio, $\gdr$, following
\begin{align}
    \Mgas = \gdr \times \Mdust.
\end{align}

We adopt the empirical relation between gas-to-dust ratio and gas-phase metallicity calibrated by \citet{2011ApJ...737...12L} and used in \citet{2018ApJ...853..179T}:
\begin{align}
    \log \gdr = \log \frac{\Mgas}{\Mdust} = 9.4 - 0.85 \times [12+\log\mathrm{(O/H)}].  
    \label{eq:GDR}
\end{align}

Gas-phase metallicities are derived from the N2 index (i.e., the [\Nii]/H$\alpha$ ratio) measured in the FMOS-COSMOS survey \citep{2014ApJ...792...75Z,2017ApJ...835...88K,2019ApJS..241...10K}, converted to oxygen abundance using the linear calibration of \citet{2004MNRAS.348L..59P}.  The resulting metallicities span $12+\log(\mathrm{O/H}) \approx 8.4$--$8.9$, with typical statistical uncertainties of $\approx 0.06$~dex.  The median metallicity of our sample is consistent with the massive end of the mass-metallicity relation at $z\sim1.5$ \citep{2014ApJ...792...75Z,2019ApJS..241...10K}. The individual measurements scatter around the established relation without a significant systematic offset, indicating that our galaxies are representative of massive MS systems at this epoch in terms of metallicity.

The corresponding gas-to-dust ratios range from $\gdr \approx 72$--178, with a median value of 100, as shown in Figure \ref{fig:Mstar_vs_GDR}.  For reference, the corresponding metallicity is indicated on the right-hand $y$-axis, and the mass-metallicity relations at $z\sim0$ and $z\sim1.5$ from \citet{2019ApJS..241...10K} are also shown.  Because our sample lies at the massive end where the mass-metallicity relation is relatively flat, $\gdr$ exhibits only weak dependence on stellar mass.  The scatter in $\log \gdr$ is $\approx 0.07$~dex, reflecting the small metallicity dispersion in this mass regime.

We note that Equation (\ref{eq:GDR}) implicitly assumes that the total cold gas mass is dominated by molecular gas.  Direct \Hi\ measurements are not available at these redshifts, and theoretical as well as indirect observational arguments suggest that atomic gas may contribute non-negligibly at $z \sim 1.5$ \citep[e.g.,][]{2021MNRAS.502L..85M}.  Therefore, our inferred $\Mgas$ may formally represent the total cold gas mass (\Hi\ + H$_2$).  In the following, we refer to it as molecular gas mass for consistency with previous scaling studies, while acknowledging this caveat.

The derived gas masses span $\Mgas \sim 10^{9.9}$--$10^{11.5}~M_\odot$, with a median of $10^{10.87}~M_\odot$ and a typical uncertainty of $\sim0.17$~dex. These values correspond to molecular gas mass ratios $\mugas \equiv \Mgas/M_\ast$ ranging from 0.14 to 3.3, with a median $\mu_{\rm gas}=0.77$.

We adopt a metallicity-dependent gas-to-dust ratio to account for galaxy-to-galaxy variations in chemical enrichment.  Although the metallicity range at the massive end is relatively narrow, this approach minimizes potential systematic biases compared to assuming a fixed $\gdr$.  Incorporating the intrinsic scatter in the N2-metallicity calibration ($\sim 0.16$~dex in $12+\log(\mathrm{O/H})$; \citealt{2013A&A...559A.114M}) does not alter our main conclusions within the quoted uncertainties.

The robust determination of $\Mgas$ for individual galaxies enables us to investigate how gas fraction and depletion time vary across the MS, as discussed in the following subsection.

\subsection{Main-sequence offset and gas scaling relations}
\label{sec:MS_offset}

To investigate the physical origin of the MS scatter, we examine how molecular gas mass ratio and depletion time vary as a function of the offset from the MS.

We adopt the MS prescription used in \citet{2018ApJ...853..179T}, based on the formulation of \citet{2014ApJS..214...15S}.  Figure~\ref{fig:Mstar_vs_SFR} shows the MS relation evaluated at the median redshift of our sample ($z=1.53$).  For each galaxy, the MS expectation is computed using its individual stellar mass and redshift.  We define the offset from the MS as $\dMS \log \sSFR = \log \left( \sSFR / \MS{\sSFR} \right)$, which is equivalent to $\log(\SFR/\MS{\SFR})$.  For the final sample of 53 galaxies, the distribution of $\dMS \log \sSFR$ has an intrinsic scatter of 0.22~dex and a mean offset of $+0.11$~dex relative to the adopted MS prescription, indicating that our sample is, on average, slightly biased toward higher SFRs than the reference MS (see Section \ref{sec:results}).

To quantify variations in gas-related quantities across the MS, we normalize them by their MS expectations at fixed stellar mass and redshift.  For the molecular gas mass ratio, $\mugas \equiv \Mgas/M_\ast$, we adopt the unified scaling relation from 
\citet{2018ApJ...853..179T}:
\begin{align}
    \log \mugas &= 0.07-3.8\times(\log(1+z)-0.63)^2 \nonumber \\
    & +0.53 \times \log~(\sSFR/\MS{\sSFR}) \nonumber \\
    & - 0.33 \times (\log M_\ast-10.7).
    \label{eq:Tacconi18}
\end{align}
The MS expectation of $\mugas$ is obtained by setting $\log~(\sSFR/\MS{\sSFR})=0$.  Adding $\log M_\ast$ to both sides gives the corresponding relation for $\Mgas$, such that 
\begin{align}
\log \left( \frac{\Mgas}{\MS{\Mgas}} \right) = \log \left( \frac{\mu_{\rm gas}}{\MS{\mu_{\rm gas}}} \right).
\end{align}

We define the MS depletion time as 
\begin{align}
    \MS{\taudep} = \frac{\MS{\Mgas}}{\MS{\SFR}}
\end{align}
and therefore the normalized depletion time can be written as 
\begin{align}
\log \left( \frac{\tau_{\rm dep}}{\MS{\tau_{\rm dep}}} \right)
=
\log \left( \frac{\Mgas}{\MS{\Mgas}} \right)
-
\log \left( \frac{\SFR}{\MS{\SFR}} \right).
\end{align}

This formulation highlights a direct mathematical link between gas mass ratio and depletion time.  If the offset in $\mugas$ scales as $(\sSFR/\MS{\sSFR})^B$, the offset in $\taudep$ scales as $(\sSFR/\MS{\sSFR})^{B-1}$. 

In Section~\ref{sec:results_dMS}, we determine the power-law index $B$ for our homogeneous sample of massive MS galaxies at $z\sim 1.5$ using individually measured gas masses.  Throughout the remainder of this paper, $\Delta_\mathrm{MS} \log X$ denotes the logarithmic offset of quantity $X$ from its MS expectation at fixed stellar mass and redshift.

\section{Results and discussion} \label{sec:results}

We present the dust and molecular gas properties of the FMOS-ALMA galaxies and examine their connection to star formation activity and the scatter around the MS.  Our primary goal is to isolate the physical drivers of star formation variability in massive galaxies at $z \sim 1.5$.

In the following analysis, all scaling relations are derived using orthogonal distance regression (ODR), accounting for uncertainties in both variables.

\subsection{Dust Content and Radiation Field in Massive Galaxies} \label{sec:results_dust}

\begin{figure}[t]
\begin{center}
\includegraphics[width=3.4in]{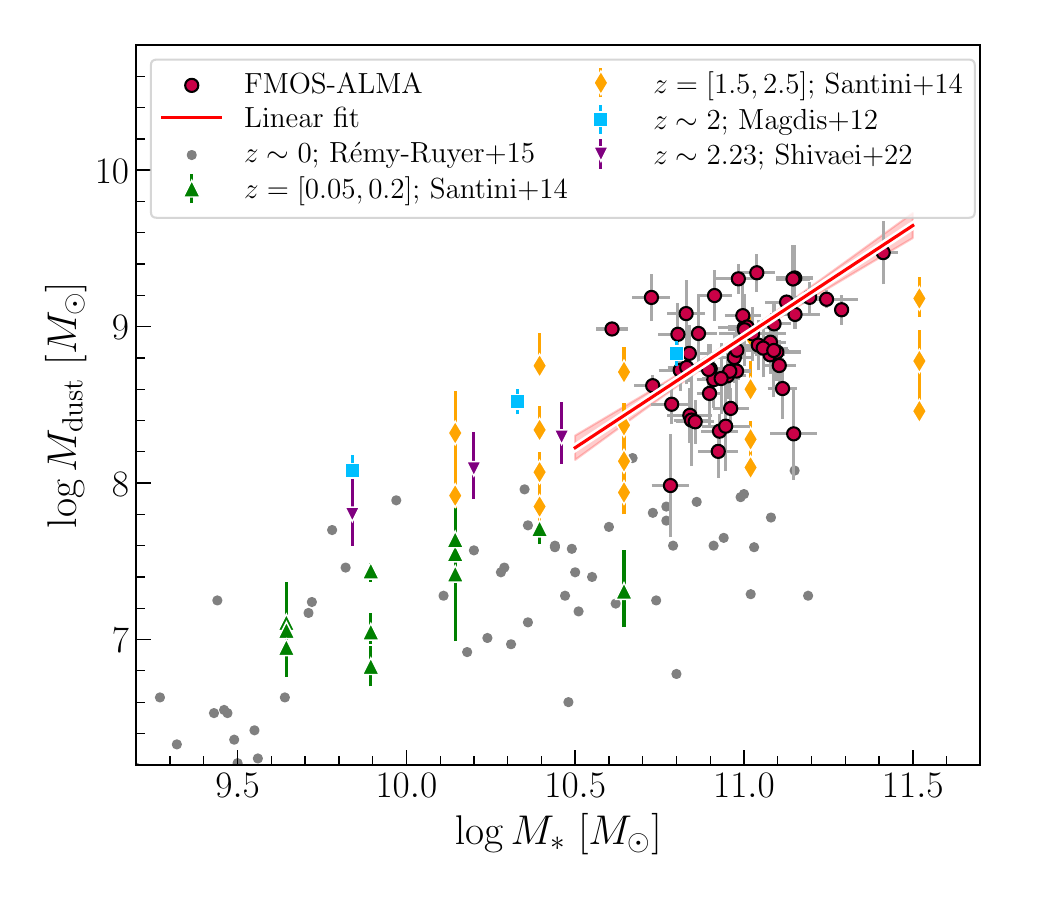}
\end{center}
\caption{Dust mass as a function of stellar mass.  The red solid line shows the best-fit relation to the FMOS-ALMA sample (Equation \ref{eq:Ms_Md_fit}).  The light red shaded region represents the $1\sigma$ confidence interval of the fit.  Measurements at $z\sim0$ and $z\sim2$ from the literature are overplotted for comparison, as indicated in the legend \citep{2012ApJ...760....6M,2014A&A...562A..30S,2015A&A...582A.121R,2022ApJ...928...68S}.
\label{fig:Mstar_vs_Mdust}}
\end{figure}

Dust masses are derived from SED fitting with the \citet{2007ApJ...657..810D} model of dust emission for all FMOS-ALMA galaxies.  Figure \ref{fig:Mstar_vs_Mdust} shows a clear correlation (Spearman's rank correlation coefficient $\rho=0.45$ with $p=6.5\times10^{-4}$) between dust mass and stellar mass.  The ODR linear fit finds the best-fit relation 
\begin{align}
    & \log\left(\frac{\Mdust}{M_\odot}\right) \nonumber \\
    & = (7.52\pm0.29) + (1.42\pm 0.29) \log \left( \frac{M_\ast}{10^{10}~M_\odot} \right).
    \label{eq:Ms_Md_fit}
\end{align}
with an intrinsic scatter of 0.20~dex.

For comparison, local measurements from \citet{2015A&A...582A.121R} and stacked $z\sim2$ measurements \citep{2012ApJ...760....6M,2014A&A...562A..30S,2022ApJ...928...68S} are shown in Figure~\ref{fig:Mstar_vs_Mdust}.
At fixed stellar mass, dust masses increase by a factor of $\sim 10$--20 from $z \sim 0$ to $z \sim 1.5$--2, consistent with the evolution observed in other studies at similar epochs.  This reflects the substantially larger gas reservoirs in galaxies at cosmic noon (Section \ref{sec:results_Mgas}). 

\begin{figure}[t]
\begin{center}
\includegraphics[width=3.4in]{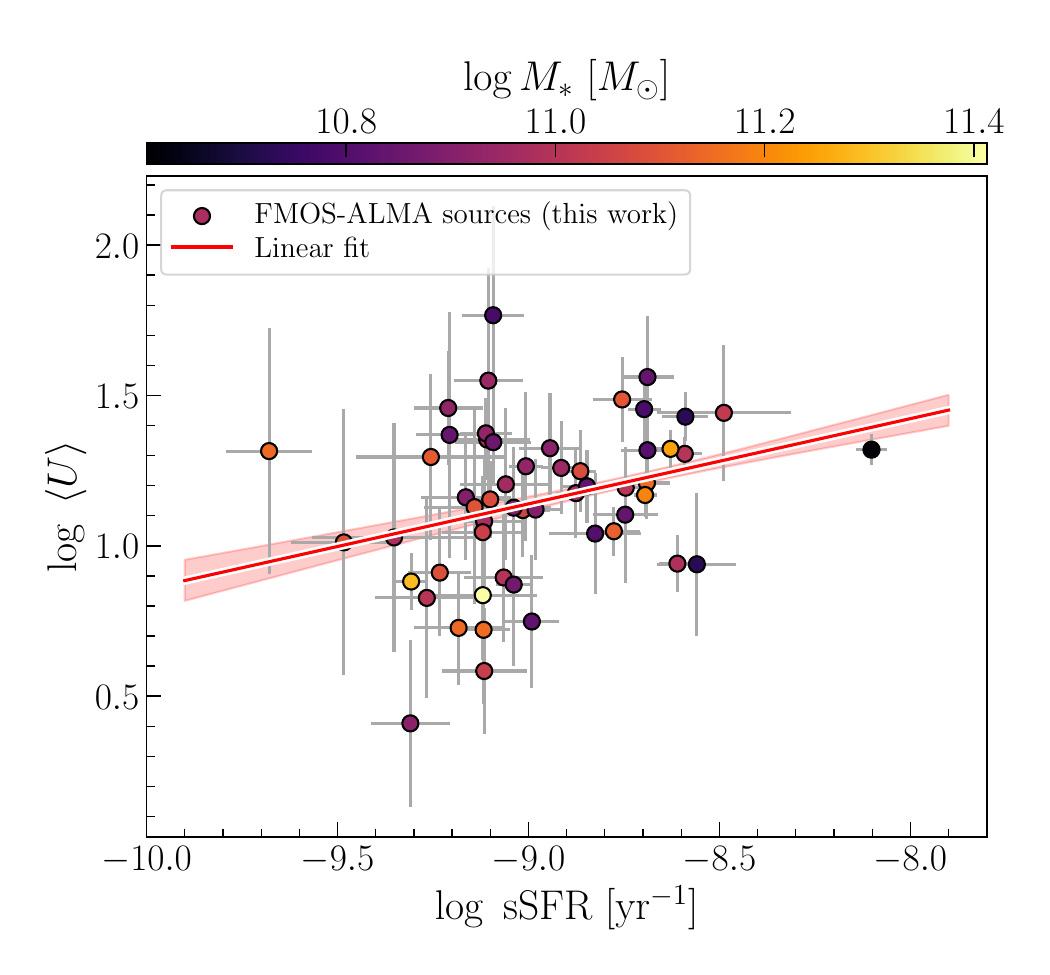}
\end{center}
\caption{Mean radiation field, $\Umean$, as a function of specific star formation rate (sSFR). Large circles are color-coded by stellar mass. The red solid line shows the best-fit relation to the FMOS-ALMA sample, with the shaded region indicating the $1\sigma$ confidence interval.
\label{fig:sSFR_vs_Umean}}
\end{figure}

We also derive the mean intensity of the stellar radiation field illuminating the dust, $\Umean \sim (\LIR/L_\odot)/(\Mdust/M_\odot)$.  This is linked to the effective dust temperature, approximately, via $\Tdust~[\mathrm{K}]\sim 18 \Umean^{1/6}$ \citep{2014ApJ...780..172D,2018A&A...609A..30S}.  The median value of our sample is $15.0\pm1.3$, corresponding to $\Tdust \approx 28~\mathrm{K}$.  This is higher than typical values for local MS galaxies, and is consistent with past measurements at similar epochs \citep{2012ApJ...760....6M,2014A&A...561A..86M,2018A&A...609A..30S}.

Figure~\ref{fig:sSFR_vs_Umean} shows $\Umean$ as a function of sSFR.  A moderate positive correlation is found between $\log \Umean$ and $\log \sSFR$, characterized by Spearman's $\rho = 0.35$ ($p=0.011$).  Given the tight correlation between $\Mdust$ and $M_\ast$ and considering $\LIR$ scales approximately with SFR, a positive correlation between these quantities is naturally expected.  Our linear regression yields
\begin{align}
    & \log \Umean \nonumber \\
    & = (1.139 \pm 0.030) + (0.283 \pm 0.075) \log \left( \frac{\sSFR}{10^{-9}~\mathrm{yr^{-1}}} \right).
    \label{eq:sSFR_umean_Md_fit}
\end{align}
The intrinsic scatter in $\log \Umean$ around this relation is estimated to be 0.12~dex.

\subsection{Molecular gas mass and comparison with scaling relations}
\label{sec:results_Mgas}

\begin{figure}[t]
\begin{center}
\includegraphics[width=3.4in]{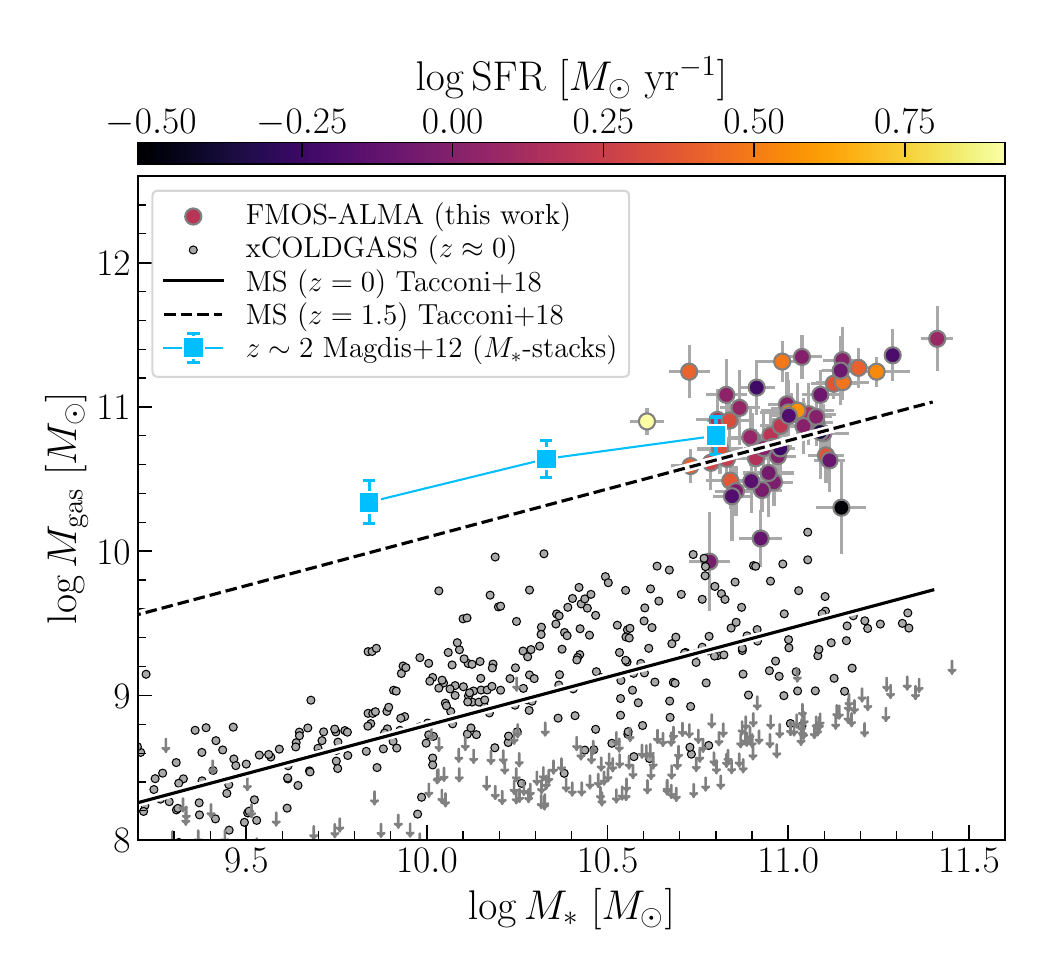}
\end{center}
\caption{Molecular gas mass as a function of stellar mass.  For the FMOS-ALMA sample, molecular gas masses are derived from dust masses using a metallicity-dependent $\gdr$ (large circles, color-coded by SED-based SFR).  Small circles represent local measurements from xCOLD GASS \citep{2017ApJS..233...22S}, with downward arrows indicating 3$\sigma$ upper limits.  Blue squares show stacked measurements for $z\sim2$ star-forming galaxies from \citet{2012ApJ...760....6M}.  The solid and dashed lines show the predictions of the unified scaling framework from \citet{2018ApJ...853..179T} at the relevant redshifts.
\label{fig:Mstar_vs_Mgas}}
\end{figure}

Figure \ref{fig:Mstar_vs_Mgas} shows the molecular gas mass as a function of stellar mass and compares our measurements with local galaxies from the xCOLD GASS survey \citep{2017ApJS..233...22S} and stacked $z\sim2$ measurements \citep{2012ApJ...760....6M}.  
At fixed stellar mass, our galaxies exhibit molecular gas contents more than an order of magnitude higher than local systems and comparable to $z \sim 2$ populations.

We further compare our sample with the empirical unified relation \citep{2018ApJ...853..179T}, which is based on a compilation of molecular gas measurements across a broad range of redshifts and tracers (see Equation \ref{eq:Tacconi18}).   The overall distribution of our galaxies is consistent with the expected relation at $z \sim 1.5$, indicating that the global gas content of massive MS galaxies follows the redshift evolution described by this framework.  Our homogeneous sample within a narrow redshift interval provides a controlled test of the scaling relation in the massive regime.

\begin{figure}[t]
\begin{center}
\includegraphics[width=3.4in]{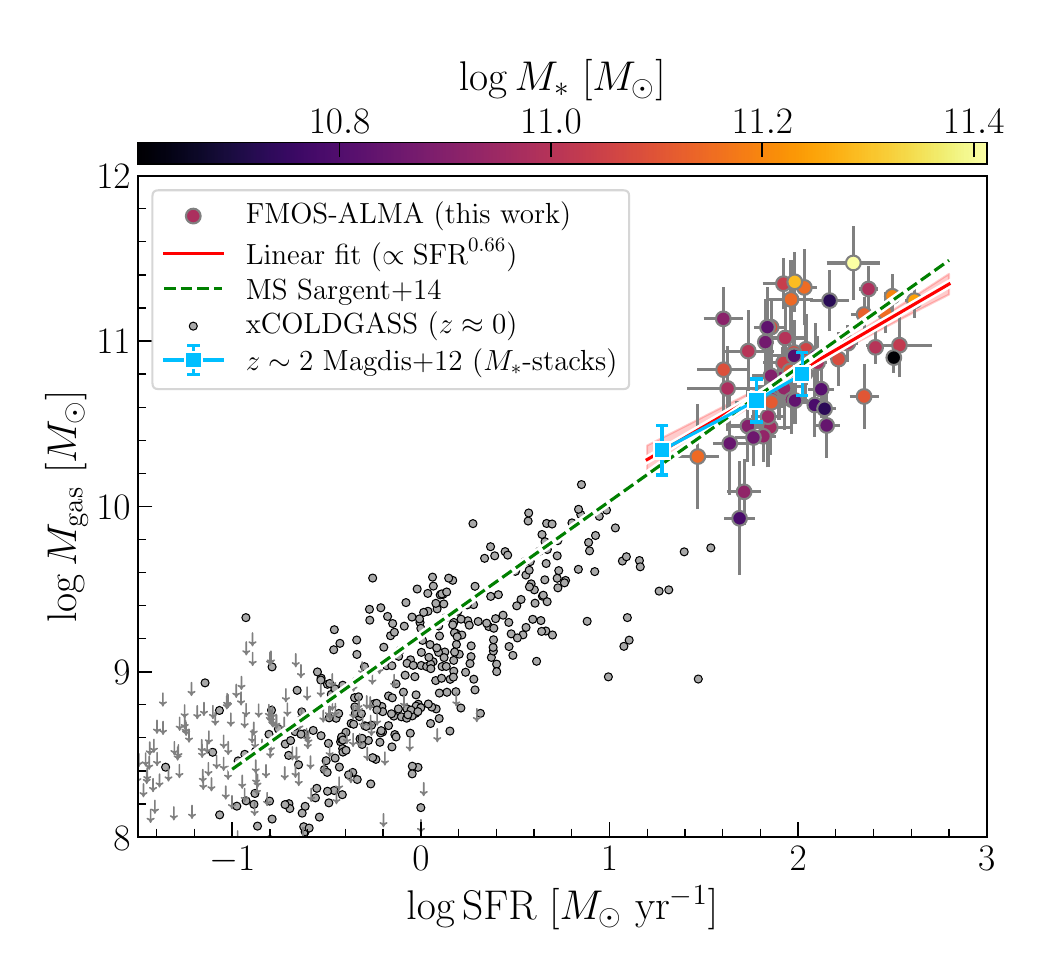}
\end{center}
\caption{Molecular gas mass as a function of SFR.  For the FMOS-ALMA sample, SFRs derived from CIGALE SED fitting are adopted (large circles, color-coded by stellar mass). The solid red line shows the best-fit linear relation.  The green dashed line indicates the main-sequence integrated Schmidt-Kennicutt relation derived by \citet{2014ApJ...793...19S}, which is in good agreement with our best-fit relation within the uncertainties.  Other symbols show the same comparison samples as in Figure \ref{fig:Mstar_vs_Mgas}.
\label{fig:SFR_vs_Mgas}}
\end{figure}

Figure \ref{fig:SFR_vs_Mgas} presents the molecular gas mass as a function of SFR.  Our sample exhibits a clear positive correlation between these quantities (Spearman's $\rho=0.48$, $p=2.4\times10^{-4}$). A linear regression yields the integrated Schmidt-Kennicutt (S-K) relation: 
\begin{align}
    &\log \left( \frac{\Mgas}{M_\odot}\right) \nonumber \\
    &= (9.49 \pm 0.27) + (0.66 \pm 0.13)\log \left( \frac{\mathrm{SFR}}{M_\odot~\mathrm{yr^{-1}}} \right).
    \label{eq:fit_SFR_vs_Mgas}
\end{align}
The intrinsic scatter in $\log \Mgas$ is estimated to be 0.20~dex.  

For comparison, we overplot the empirical relation derived by \citet{2014ApJ...793...19S} for MS galaxies, based on compiled measurements over $z=0$--2.  Our best-fit relation is consistent, within the uncertainties, indicating that massive MS galaxies at $z\sim1.5$ obey the same global S-K relation observed over a broad redshift range.  This result provides a consistent baseline for interpreting variations around the MS in terms of gas mass ratio and depletion time.

\subsection{Decomposing the main sequence scatter}
\label{sec:results_dMS}

\begin{figure*}[t]
\begin{center}
\includegraphics[width=3.4in]{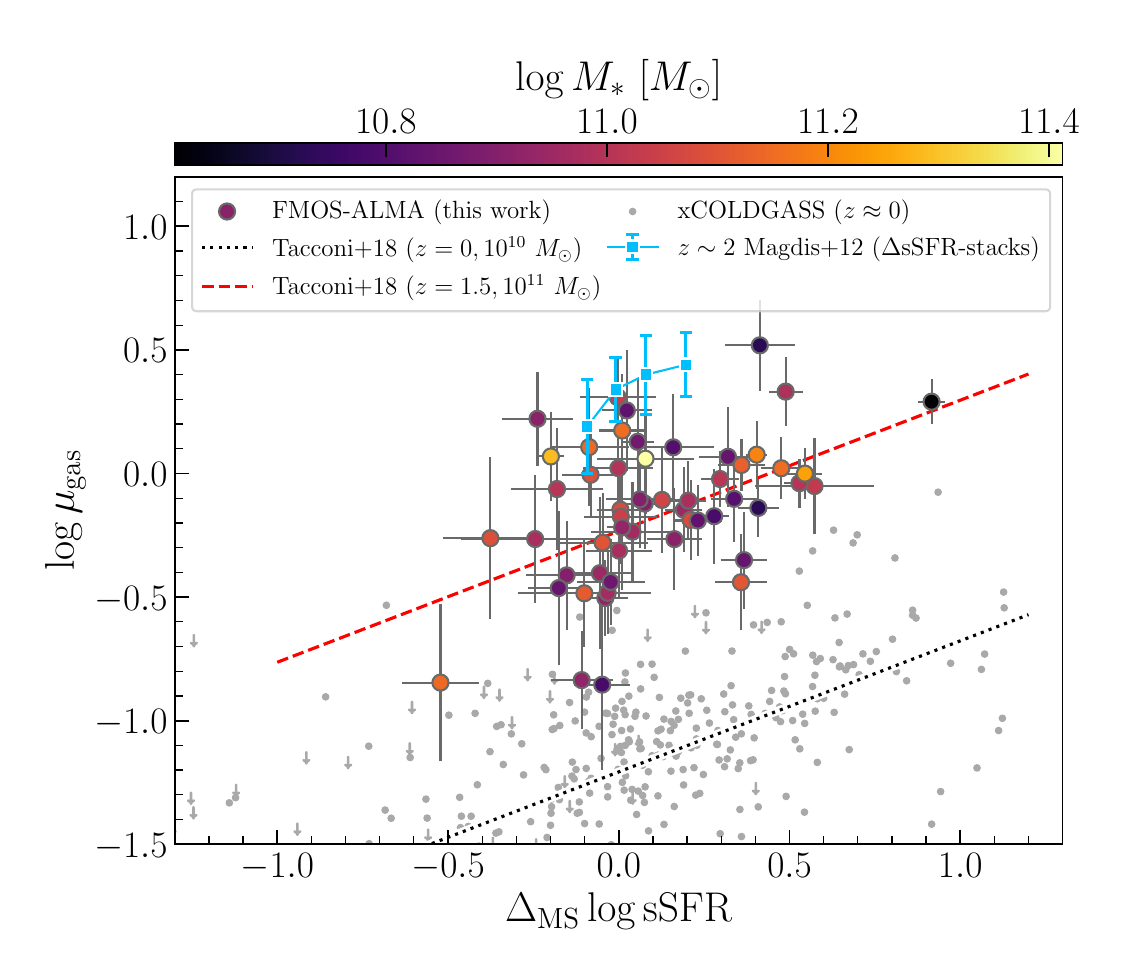}
\includegraphics[width=3.4in]{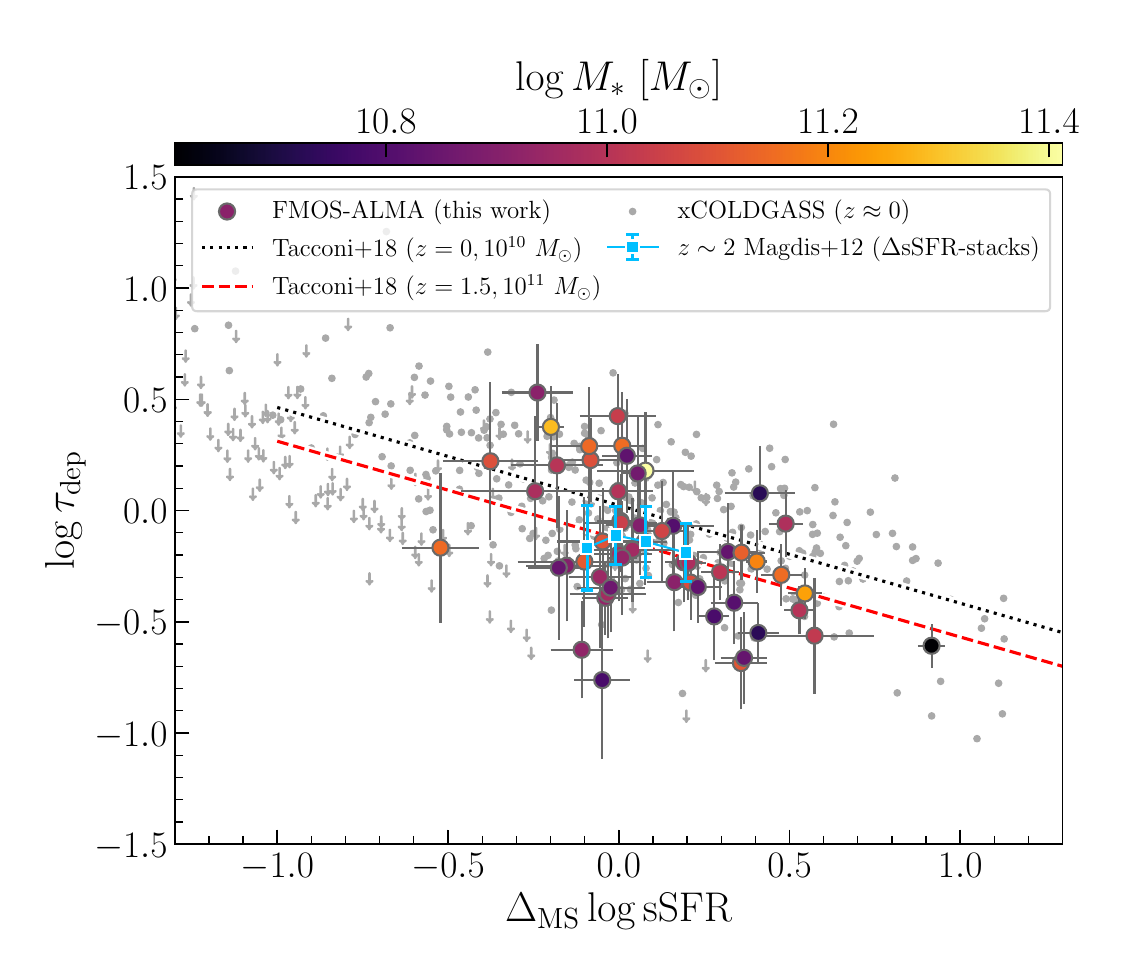}
\end{center}
\caption{Gas mass ratio $\mugas = \Mgas/M_\ast$ (left panel), and depletion time $\taudep = \Mgas/\mathrm{SFR}$ (right panel), as a function of MS-normalized sSFR.  Large circles show our FMOS-ALMA sources, color-coded by stellar mass.  The black dotted line and red dashed lines indicate the unified empirical scaling relations of \citet{2018ApJ...853..179T} at $z=0$, $M_\ast = 10^{10}~M_\odot$, and at $z=1.5$, $M_\ast = 10^{11}~M_\odot$, respectively.  The latter is chosen to match the median stellar mass of our sample.  Gray dots show local xCOLD GASS measurements \citep{2017ApJS..233...22S}, with downward arrows indicating $3\sigma$ upper limits, and blue squares show stacked measurements at $z\sim2$ from \citet{2012ApJ...758L...9M}.
\label{fig:dMS_mugas_tdep}}
\end{figure*}
\begin{figure*}[t]
\begin{center}
\includegraphics[width=3.4in]{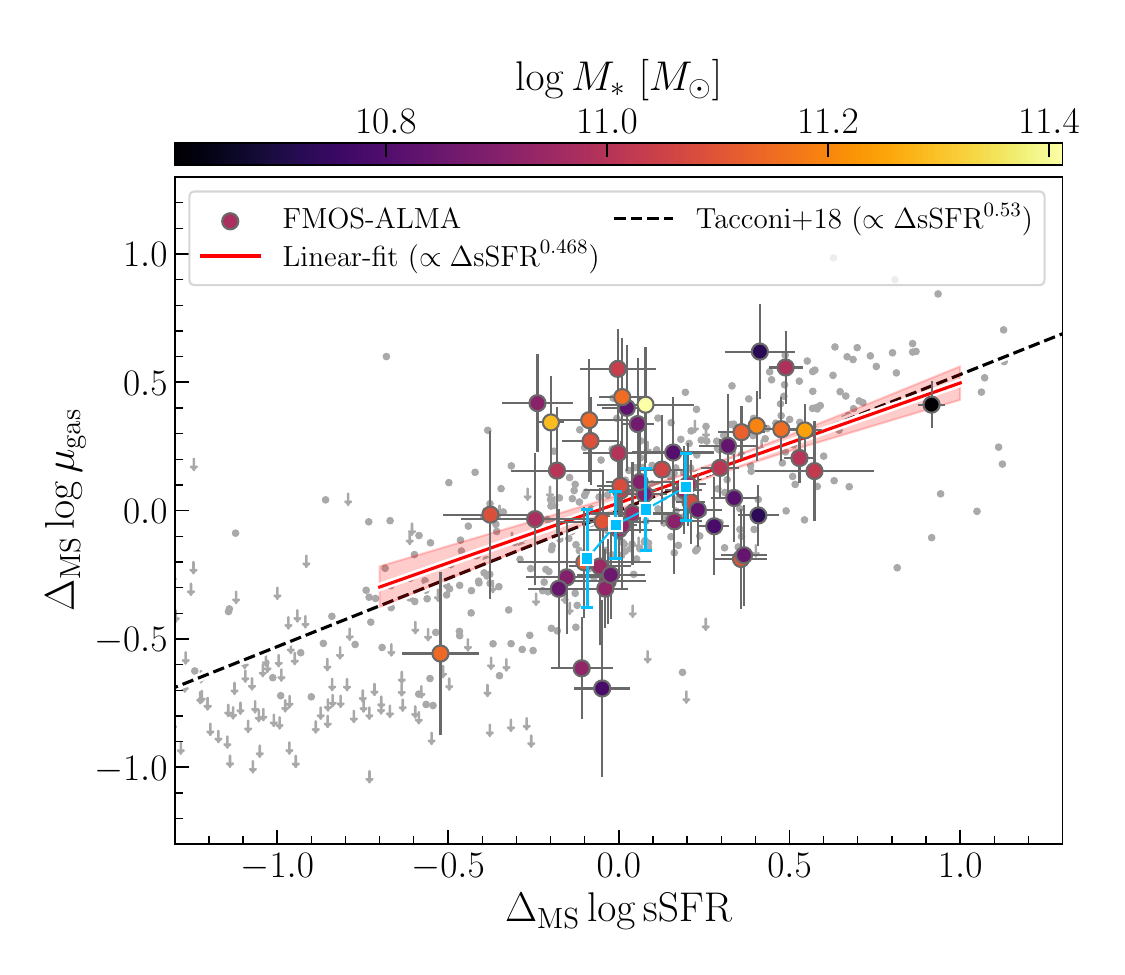}
\includegraphics[width=3.4in]{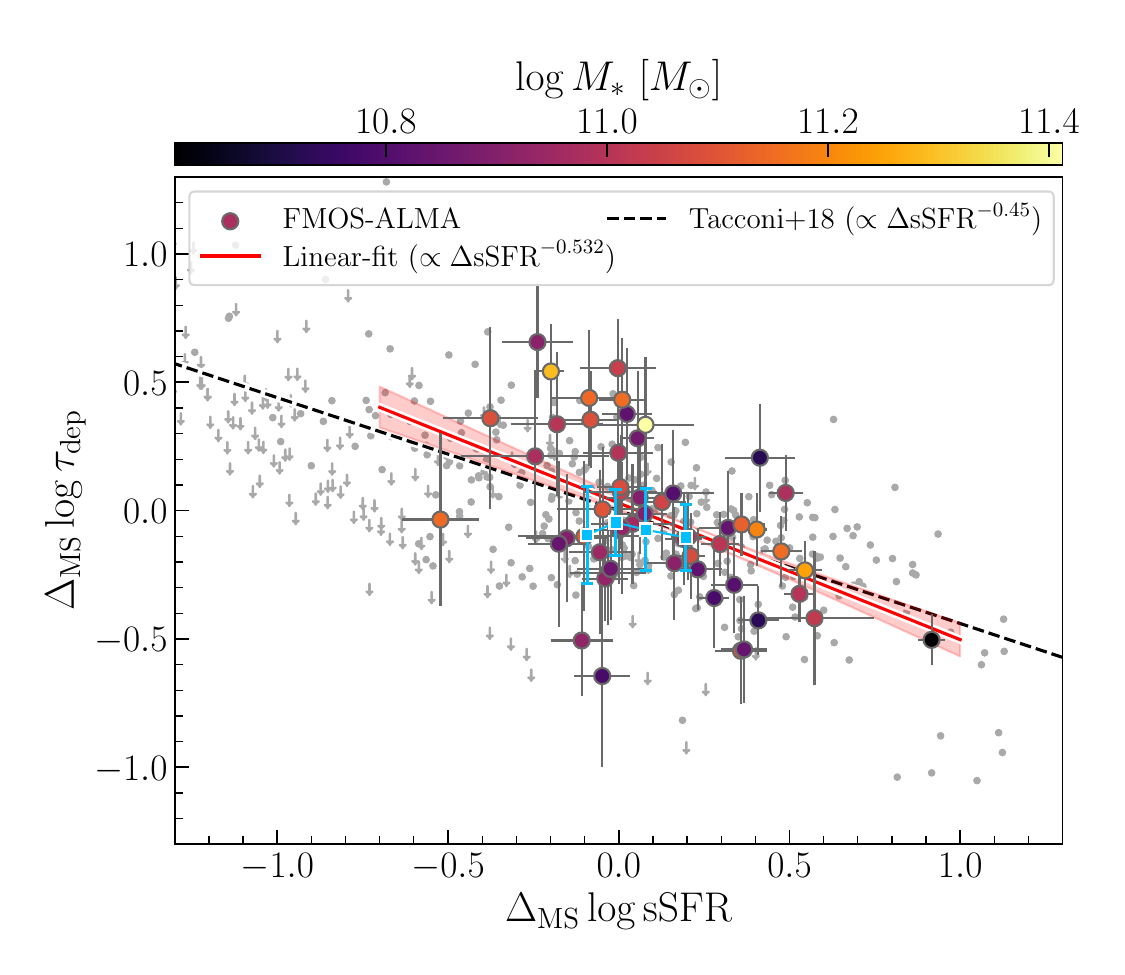}
\end{center}
\caption{MS-normalized gas mass ratio (left panel), and depletion time (right panel), as a function of MS-normalized sSFR (large circles, color-coded by stellar mass).  The red solid line shows the best-fit linear relation to the FMOS-ALMA sample in each panel.  The black dashed line indicate the unified empirical scaling relations from \citet{2018ApJ...853..179T}.  The literature data points are the same as in Figure \ref{fig:dMS_mugas_tdep}.
\label{fig:dMS}}
\end{figure*}

We now investigate the physical origin of the scatter around the main sequence.  
Figure \ref{fig:dMS_mugas_tdep} shows the gas mass ratio, $\mugas$, and depletion time, $\taudep$, as a function of the offset from the MS, $\dMS \log \sSFR$.  For reference, we also plot the unified scaling relations of \citet{2018ApJ...853..179T} at $z=0$, $M_\ast = 10^{10}~M_\odot$, and at $z=1.5$, $M_\ast=10^{11}~M_\odot$.  The latter is chosen to match the median stellar mass of our sample.  Local measurements from xCOLD GASS \citep{2017ApJS..233...22S} and stacked measurements at $z\sim2$ from \citet{2012ApJ...758L...9M} are also shown.  These comparisons clearly show that the gas mass ratios in our FMOS-ALMA sources at $z\sim1.5$ are significantly higher than those of MS galaxies at $z\sim0$, whereas the depletion times show little evolution.  This behavior is consistent with the unified scaling relation, and suggests that the overall redshift evolution of the MS is driven primarily by the evolution of gas content rather than by changes in depletion time.

We next focus on decomposing the scatter around the MS at a single epoch, $z\sim 1.5$, using our sample.  All quantities are now normalized relative to their MS expectation at fixed stellar mass, following the prescription described in Section \ref{sec:MS_offset}.  Figure \ref{fig:dMS} shows that both the MS-normalized molecular gas mass ratio, $\mugas~(=\Mgas/M_\ast)$, and depletion time, $\taudep$, correlate with the offset from the MS, $\dMS \log \sSFR$.  The best-fit relation for the gas mass ratio is 
\begin{align}
    & \log \frac{\mugas}{\MS{\mugas}}  = \log \frac{\Mgas}{\MS{\Mgas}} \nonumber \\
    & = (0.029\pm0.041) + (0.47 \pm 0.11) \log \frac{\sSFR}{\MS{\sSFR}},
    \label{eq:mugas}
\end{align}
with the intrinsic scatter in $\dMS \log \mugas$ estimated to be 0.16~dex.
The intercept is formally allowed to vary in the fitting procedure to account for uncertainties in the adopted MS normalization; in the ideal case of perfectly determined MS parameters, it would be zero.

This relation can be equivalently expressed in terms of the depletion time as
\begin{align}
    & \log \frac{\taudep}{\MS{\taudep}}  \nonumber \\
    & = (0.029\pm0.041) + (-0.53 \pm 0.11) \log \frac{\sSFR}{\MS{\sSFR}}.
    \label{eq:taudep}
\end{align}

The nearly symmetric slopes indicate that variations in molecular gas supply and star formation efficiency contribute comparably to the MS scatter.  At fixed stellar mass, the offset from the MS can be decomposed as $\dMS \log \mathrm{SFR} \simeq \dMS \log \mugas
+ \dMS \log \mathrm{SFE}$, where $\mathrm{SFE}=\taudep^{-1}$.  Our results therefore suggest that the variation of SFR across the MS is not driven solely by changes in the available molecular gas reservoir, but by a coupled modulation of both gas content and the efficiency of converting gas into stars.  This interpretation lies between pictures in which SFR variations are attributed primarily to gas content \citep[e.g.,][]{2016ApJ...820...83S} and those in which enhanced SFE plays a larger role for galaxies above the MS \citep[e.g.,][]{2015ApJ...812L..23S}.  These slopes are consistent with the unified scaling framework of \citet{2018ApJ...853..179T}, in which both $\mugas$ and $\taudep$ vary systematically with distance from the MS.
Because our analysis is based on a homogeneous sample within a narrow redshift interval ($1.45<z<1.70$), evolutionary mixing and methodological differences are minimized.  This provides a controlled test of the unified scaling framework in the massive regime at cosmic noon, and supports the picture in which the MS scatter is primarily driven by coupled variations in gas content and star formation efficiency.

At the same time, recent ALMA studies have highlighted the structural complexity of massive MS galaxies at similar redshifts.  In many systems, the submillimeter emission is significantly more compact than the stellar light, indicating centrally concentrated star formation \citep[e.g.,][]{2019ApJ...877L..23P,2020ApJ...901...74T,2021MNRAS.508.5217P}.  Such compact components can exhibit shorter local depletion times while the galaxy as a whole remains on the MS in terms of integrated quantities.  In this context, the global variation in $\taudep$ with MS offset found here may partly reflect changes in the prominence of compact, high-efficiency star-forming regions embedded within more extended disks.

The remaining intrinsic scatter of $\approx 0.16$~dex likely reflects a combination of residual systematic uncertainties (e.g., metallicity calibration and gas-to-dust conversion) and genuine galaxy-to-galaxy diversity in star formation efficiency.  While the precise origin of this residual dispersion remains unclear, the dominant role of gas content and depletion time indicates that the same regulation mechanism inferred from compiled datasets is already in place in massive MS galaxies at $z\sim1.5$.

\section{Summary}
\label{sec:summary}

We have carried out ALMA Band 7 observations to measure dust continuum emission (rest-frame $\sim350~\mathrm{\mu m}$) for 56 massive ($M_\ast \gtrsim 10^{10.8}~M_\odot$) star-forming galaxies at $1.45<z<1.70$ located along the main sequence (MS) at cosmic noon, and retrieved archival data of comparable quality for one additional source. 
Among the 57 galaxies, 55 are detected at $>3\sigma$ in dust continuum, and all sources are detected at $>2\sigma$.  
Combining these measurements with multiwavelength photometry from UV to FIR, we performed SED fitting using the \citet{2007ApJ...657..810D} dust emission model.  After excluding two AGN candidates and one source with an unreliable SED fit, dust masses were derived for the final statistical sample and converted to molecular gas masses using metallicity-dependent gas-to-dust ratios.

Our main findings are as follows.

(1) The dust and molecular gas contents of massive star-forming galaxies at $z \sim 1.5$ are $\sim10$--20 times higher than those of local galaxies at fixed stellar mass.  The median molecular gas mass ratio $\mugas$ is 0.65.  The integrated S-K relation (SFR versus $\Mgas$) for our sample is consistent with the extrapolation from local galaxies and with previous measurements at similar epochs.

(2) Across the MS, both the molecular gas mass ratio $\mugas$ and the star formation efficiency $\mathrm{SFE}=\taudep^{-1}$ scale approximately as $(\mathrm{sSFR}/\langle \mathrm{sSFR}\rangle_\mathrm{MS})^{1/2}$.  Since, at fixed stellar mass, the MS offset can be decomposed as $\dMS \log \mathrm{SFR} \simeq \dMS \log \mugas + \dMS \log \mathrm{SFE}$, this implies that the scatter of the MS is driven almost equally by variations in gas content and depletion time.  This result is consistent with the unified scaling framework of \citet{2018ApJ...853..179T}, now tested here using a homogeneous sample within a narrow redshift interval in the massive regime.  

Overall, our results demonstrate that the physical regulation of star formation through the coupled modulation of gas content and star formation efficiency is already established in massive main-sequence galaxies at $z \sim 1.5$.  The present dataset, together with high-resolution imaging from the COSMOS-Web program \citep{2023ApJ...954...31C}, will enable spatially resolved studies of stellar structure, star formation activity, and ISM conditions.  Future work will explore how these processes connect to stellar mass growth, structural evolution, and transitions in star formation activity.



\begin{ack}
This paper makes use of the following ALMA data: ADS/JAO.ALMA\#2021.1.01133.S, ADS/JAO.ALMA\#2015.1.00568.S. ALMA is a partnership of ESO (representing its member states), NSF (USA) and NINS (Japan), together with NRC (Canada), NSTC and ASIAA (Taiwan), and KASI (Republic of Korea), in cooperation with the Republic of Chile. The Joint ALMA Observatory is operated by ESO, AUI/NRAO and NAOJ.  This research is based in part on data collected at the Subaru Telescope, which is operated by the National Astronomical Observatory of Japan. We are honored and grateful for the opportunity of observing the Universe from Maunakea, which has cultural, historical, and natural significance in Hawaii.  This work is supported by JSPS KAKENHI Grant Number JP25K07362.
\end{ack}



\appendix 
\section*{CIGALE SED fitting} \label{sec:apx_cigale}

Table \ref{tab:cigale} summarizes the full set of variable parameters and their adopted ranges in the CIGALE fitting.

\begin{table*}
  \tbl{Input parameters of the SED fitting with \texttt{CIGALE}}{%
  \small
  \begin{tabular}{ll}
      \hline
      Parameter & Values \\
      \hline
      \multicolumn{2}{c}{delayed$+$burst SFH} \\
      \hline
      $e$-folding time of the delayed SFH, $\tau_0$ [Myr] 
        & 1000, 2000, 3000 \\
      Age of the main population [Myr] 
        & 1000--4000 in steps of 500 \\ 
      $e$-folding time of the late starburst, $\tau_1$ [Myr] 
        & 50 \\
      Age of the late burst [Myr] 
        & 20 \\ 
      Mass fraction of the late burst population  
        & 0.0, 0.005, 0.010, 0.020, 0.050, 0.100 \\
      \hline
      \multicolumn{2}{c}{Dust attenuation: modified starburst law based on \citet{2000ApJ...533..682C}} \\
      \hline
      $E(B-V)_{\mathrm{line}}$ 
        & 0.227, 0.341, 0.455, 0.568, 0.682, 0.909, 1.023, 1.136, 1.25, 1.363, 1.590, 1.818 \\
      $E(B-V)_{\mathrm{star}} / E(B-V)_{\mathrm{line}}$ 
        & 0.44 \\
      UV bump wavelength [nm] \& width [nm]
        & 217.5 \& 35 \\
      UV bump amplitudes 
        & 0.0, 0.5, 1.0, 1.5, 2.0, 2.5, 3.0 \\
      Power-law slope 
        & $-0.6$, $-0.4$, $-0.2$, 0.0, 0.2, 0.4 \\
      \hline
      \multicolumn{2}{c}{Dust emission: \citet{2007ApJ...657..810D}} \\
      \hline
      Mass fraction of PAH $q_{\mathrm{PAH}}$ 
        & 0.47, 1.12, 1.77, 2.50, 3.19, 3.90, 4.58 \\
      Minimum radiation field $U_{\mathrm{min}}$ 
        & 0.1, 0.2, 0.5, 1.0, 2.0, 3.0, 5.0, 7.0, 10.0, 15.0, 20.0, 25.0 \\
      Maximum radiation field $U_{\mathrm{max}}$ 
        & $10^{6}$ \\
      Fraction illuminated by intense radiation $\gamma$ 
        & 0.0, 0.01, 0.02, 0.05, 0.1, 0.2, 0.4, 0.6, 0.8, 1.0 \\
      \hline
  \end{tabular}}
  \label{tab:cigale}
\end{table*}

\begin{figure}[t]
\begin{center}
\includegraphics[width=3.in]{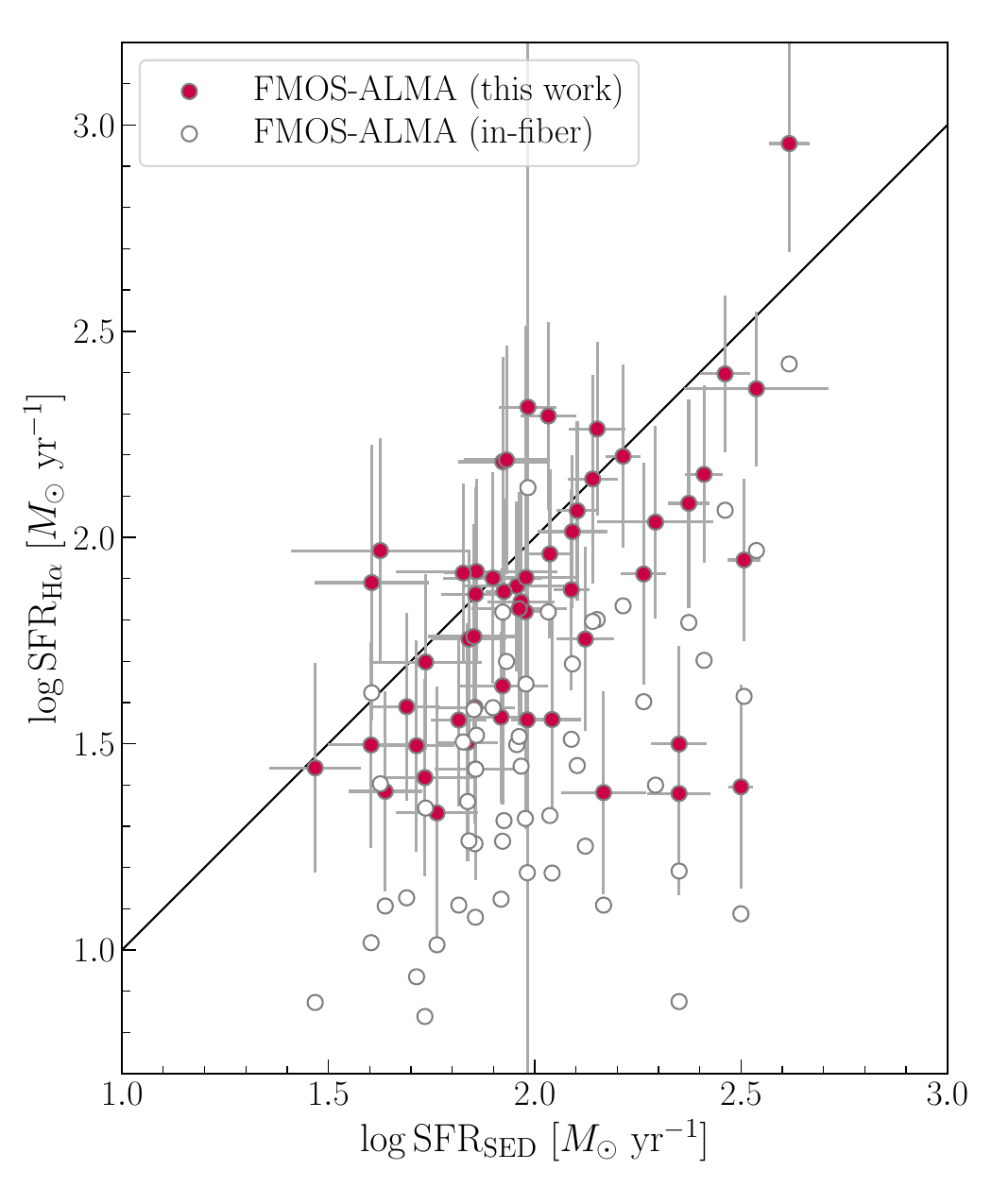}    
\end{center}
\caption{Comparison of SED-based SFRs and H$\alpha$-based SFRs. Filled circles indicate H$\alpha$-based SFRs derived from aperture-corrected FMOS H$\alpha$ luminosities, while open circles show H$\alpha$-based SFRs before aperture correction; in both cases, the H$\alpha$ luminosities are corrected for dust attenuation. The solid line represents the one-to-one relation.
\label{fig:SFRsed_vs_SFRHa}}
\end{figure}

Figure \ref{fig:SFRsed_vs_SFRHa} compares the SFRs derived from SED fitting (see Section \ref{sec:sed-fitting}) with those estimated from dust-corrected H$\alpha$ luminosities measured in the FMOS-COSMOS survey \citep{2019ApJS..241...10K}.  Before aperture correction, the H$\alpha$-based SFRs are systematically lower by a factor of $\sim 3$ on average, reflecting the limited 1.2\arcsec-diameter aperture of the FMOS fibers. After aperture correction, the two estimates are broadly consistent, with a negligible systematic offset (median $\log \mathrm{SFR_{H\alpha}}-\log \mathrm{SFR_{SED}}\approx -0.1~\mathrm{dex}$), although significant scatter remains.  This scatter likely reflects the larger systematic uncertainties in the H$\alpha$-based SFRs (see the main text).



\bibliographystyle{plainnat2}
\bibliography{ms}

\end{document}